\documentclass[preprint,11pt]{aastex}
\usepackage{natbib}
\usepackage{graphicx}
\shorttitle{The Search for Extra-solar Planets with SIM}
\shortauthors{Sozzetti et al.}

\begin{document}


\title{Narrow-Angle Astrometry with the Space Interferometry Mission: \\
       The Search for Extra-solar Planets. \\
       I. Detection and Characterization of Single Planets}


\author{A. Sozzetti\altaffilmark{1,2,4}}
\altaffiltext{1}{University of
Pittsburgh, Dept. of Physics \& Astronomy, Pittsburgh, PA 15260, USA}
\altaffiltext{2}{Smithsonian Astrophysical Observatory, Harvard-Smithsonian 
Center for Astrophysics, 60 Garden Street, Cambridge, MA 02138}
\email{alex@phyast.pitt.edu}

\author{S. Casertano\altaffilmark{3} and R. A. Brown\altaffilmark{3}}
\altaffiltext{3}{Space Telescope Science Institute, Baltimore, MD 21218,
USA}
\email{stefano@stsci.edu; rbrown@stsci.edu} 

\and

\author{M. G. Lattanzi\altaffilmark{4}}
\altaffiltext{4}{Osservatorio Astronomico di Torino, 10025 Pino Torinese,
Italy}
\email{lattanzi@to.astro.it}





\begin{abstract}
A decade after the publication of the Hipparcos Catalogue, the
Space Interferometry Mission (SIM) will be capable of making
selected high-precision astrometric measurements about three
orders of magnitude more accurate than the Hipparcos survey.

We present results from a detailed set of end-to-end
numerical simulations of SIM narrow-angle astrometric measurements
and data analysis to illustrate the enormous potential that SIM
has for the discovery and characterization of planets outside the
Solar System. Utilizing a template observing scenario, 
we quantify SIM sensitivity to single planets
orbiting single normal nearby stars as function of measurement
errors and properties of the planet:
SIM will detect over 95\% of the planets with
periods between a few days and the 5-year nominal mission lifetime
that produce astrometric signatures $\sim 2.2$ times larger than
the single-measurement accuracy. We provide
accuracy estimates of full-orbit reconstruction and planet mass
determination: at twice the discovery limit, orbital elements will
be determined with a typical accuracy of 20-30\%; the astrometric
signature must be $\sim 10$ and $\sim 15$ times the
minimum signal required for detection to derive mass and
inclination angle estimates accurate to 10\%.
We quantify the impact of different
observing strategies on the boundaries for secure detection
and accurate orbit estimation: the results scale 
with the square root of both the number of observations and the 
number of reference stars. We investigate SIM discovery space, 
to gauge the instrument ability in detecting very low-mass planets: 
around the nearest stars, SIM
will find planets as small as Earth, if they are present. Some of
these might be orbiting inside the parent star's Habitable Zone.

Extra-solar planets figure prominently among SIM scientific goals:
our results reaffirm the importance of high-precision
astrometric measurements as a unique complement to spectroscopic
surveys based on radial velocity.
For example, establishing the existence of rocky, perhaps
habitable planets would constitute both a fundamental test of
theoretical models, and progress towards the understanding of formation
and evolution processes of planetary systems. Such discoveries
would also provide the Terrestrial Planet Finder (TPF) with prime
targets to investigate with direct spectroscopy in terms of the potential
for life.

\end{abstract}



\section{Introduction}

Six years ago, research in planetary science was essentially synonymous 
with studies of our Solar System alone. Only four years later, thanks to
extensive precision radial velocity surveys of the solar neighborhood, 
Butler et al.~\cite{butler00} could list 50 nearby Main-Sequence stars 
orbited by at least one 
planet candidate with projected masses\footnote{Radial
velocity techniques cannot determine the viewing geometry of
the orbit, and consequently only lower limits to the companion
mass can be inferred} below the so-called deuterium burning
threshold ($M\sin i < 13 M_\mathrm{J}$, where $M_\mathrm{J}$ is 
the mass of Jupiter), as discussed by Oppenheimer et al.~\cite{oppen00}. 
The number of planets continues to grow; as of July 2002, 
26 more candidate planets have been identified, bringing the number
of stars harboring planetary-mass companions to 88. 
Recently, one of these candidate extra-solar planets was
confirmed to be a Jupiter-mass object in a few-days period orbit (a {\it
Hot Jupiter}) via transit observations of the star 
HD 209458~\citep{henry00,charbon00}, and the presence of 
sodium in its atmosphere detected~\citep{charbon02}. Furthermore, radial
velocity measurements have also proven the existence of candidate
planetary {\it systems}~\citep{butler99,marcy,udry,fischer02}, 
and of systems composed of a planet and a brown dwarf 
candidate~\citep{udry,marcy2,els}. 
Except for the case of HD 209458, all low-mass
companions to solar-type stars having $M\sin i < 13 M_\mathrm{J}$
have been classified by some as extra-solar planets solely on the basis of
their small projected masses, and thus, under the reasonable
assumption of orbital planes randomly oriented in space, small
true masses. In fact, the true nature of such objects is still
matter of ongoing debates among the scientific community. For
example, the unexpected orbital configurations of the majority of
the planet candidates, such as companions having $M\sin i\geq
M_\mathrm{J}$ and orbital periods of a few days~\citep{mayor95,butler97} 
or large eccentric orbits~\citep{mazeh96,coch97} have raised crucial
questions about their origin. In response to these challenging
discoveries, new theoretical models have been proposed, which
invoke diverse mechanisms like orbital 
migration~\citep{murray98,trilling98,delpopolo02} or {\it in situ} 
formation~\citep{ward97,wuch97,bode00}. 
Statistical analyses have also been carried 
out~\citep{heac99,step00,mayor00,step01,mazeh01}, which, 
highlighting the striking
similarity between the distributions of eccentricities and periods
of the two populations, suggest alternative scenarios implying
common formation processes for planet and brown dwarf candidates, 
and for stellar binaries. On the other hand, recent attempts to
determine the actual mass distribution of low-mass companions to
nearby stars~\citep{mazeh01,jorissen01,zucker01b,halbwachs01} have 
confirmed the indications obtained by early 
studies~\citep{basri97,mayor98a,mayor98b,marcy98} 
which pointed out remarkable differences in the mass distribution
of planet candidates and low-mass stellar secondaries. In particular, the
two populations appear to be separated by a gap in mass of roughly
an order of magnitude in the range 10-100 $M_\mathrm{J}$. This is
the so-called ``brown-dwarf desert'' (see for example the early
works of Campbell et al.~\cite{campbell88}, 
Marcy \& Benitz~\cite{Marcy89}, Marcy \& Butler~\cite{marcy94}, 
and more recently Halbwachs et al.~\cite{halbwachs00}), commonly
thought of as supporting the idea that the two populations are
actually distinct, and therefore suggesting that planet candidates
are indeed planets. On the basis of joint analyses of Hipparcos
Intermediate Astrometric Data and ground-based astrometric observations,
Gatewood et al.~\cite{gatewood01} and Han et al.~\cite{han01} 
cast doubts on the 
actual planetary nature of the low-mass objects detected by radial
velocity surveys, disputing the hypothesis of randomness of the
orbital planes. Recently, Pourbaix~\cite{pourbaix01a}, Pourbaix \& 
Arenou~\cite{pourbaix01b}, and Zucker \& Mazeh~\cite{zucker01a} 
have questioned the statistical
significance and robustness of their method and results, arguing that
the present milli-arcsecond precision of the most accurate astrometric
measurements today available is insufficient to derive sensible 
conclusions on the exact nature of these objects. 
Furthermore, the Main-Sequence stars harboring the planet candidates
have been shown to have higher metallicity than the average of field
stars with the same mass in the solar 
neighborhood~\citep{laughlin00,gonzalez01,santos01}, and these findings may 
support the evidence for significant
correlation between high stellar metallicity and the presence
of orbiting giant planets, under the assumption that core accretion 
is the primary planet formation mechanism. 
However, if disk instability is the preferred mechanism for forming 
extrasolar giant planets, then the metallicity dependence may 
turn out to be an artifact due to observational selection effects or 
stellar pollution by ingestion of planetary material, and even 
low-metallicity stars should harbor giant planets~\citep{boss02}. 
As it can be easily understood, such a
plethora of diverse interpretations clearly indicates how our present
understanding of the origin of planetary systems is {\it de facto}
still limited, and significant contributions in terms of data obtained
via means other than Doppler shift measurements are essential in order
to be able to discriminate between biased theoretical and
observational models.

Radial velocity surveys, accurate to 3-5 m/s~\citep{butler96}, 
have been so far a unique tool for planet discovery. However, we
anticipate that high-precision astrometry, both from 
ground \citep{mariotti98,booth99,colavita99} and in 
space \citep{danner99,roser99,gilm00}, will be among the
preferred means for helping fill regions of the parameter space
Doppler techniques cannot reach. Astrometry has a significant
advantage over the radial velocity technique because it measures
two rather than one projection of an orbit and thus describes the
full three-dimensional geometry. Astrometry, removing the degeneracy
on the inclination angle, provides unambiguous mass estimates and
directly determines coplanarity for systems of planets.
Astrometric techniques can be used to search for planets around
young and bright stars (earlier than F), and late M dwarfs. These
objects cannot be searched by radial velocity, either because of
their spectral properties (absence of relevant spectral lines) or
because of intrinsic instability of the stellar atmospheres
(active chromospheres, spots, significant rotation). Furthermore,
astrometric sensitivity increases for planets with longer periods,
thus complementing radial velocity searches, which favor
short-period planets. Finally, radial velocity detection limits
are currently of about a Saturn mass within 1 AU. As we will see,
astrometry with SIM's exquisite sensitivity pushes detection 
two orders of magnitude lower, down to Earth masses.

SIM (Space Interferometry Mission) is under development as NASA's
first space-based optical interferometer devoted to
micro-arcsecond ($\mu$as) astrometry~\citep{danner99}. It
represents a First Generation Mission within NASA's {\it Origins}
Program (\url{http://origins.jpl.nasa.gov/}), which has the long-term goal
of direct imaging of Earth-like planets around nearby solar-type
stars. The instrument is scheduled for launch by mid-2009, with a
nominal mission lifetime of 5 years. SIM will perform pointed 
observations, unlike astrometric missions such as Hipparcos 
(\url{http://astro.estec.esa.nl/Hipparcos/}), DIVA~\citep{roser99}, 
or the recently approved ESA Cornerstone Mission GAIA~\citep{perryman01}, 
which are designed to survey the sky using a well-defined scanning law, 
in order to build global astrometric catalogues. On one
hand, this will limit the total mission throughput. A few tens of thousands
objects will be observed, compared to the 120\,000 stars
surveyed by Hipparcos or to the $10^7-10^9$ objects which are expected
to be charted by DIVA and GAIA, respectively.
On the other hand, SIM's pointed observations can achieve unprecedented
astrometric accuracy that will bring new light in the
exploration of our galactic neighborhood.

Detection and measurement of planets will be carried out primarily
with SIM operated in narrow-angle astrometric mode. The instrument is
expected to achieve a narrow-angle {\it single
measurement} accuracy of $\sim 1$ $\mu$as in 1 hr integration
time on bright targets ($V\leq 11$), which corresponds to the 
amplitude of the gravitational perturbation induced on a 
solar-mass star by an Earth-mass planet on a 1 AU orbit, as seen from 
3 pc. SIM local astrometry is therefore uniquely suited for detection
and measurement of planets with masses as small as a few Earth
masses in the vicinity of the Solar System.

This is the first of two papers which will connect and relate 
the basic SIM capabilities to
the properties of extra-solar planetary system. We have built a
detailed software suite to $a)$ simulate sample narrow-angle SIM
observing campaigns of stars with planets, and $b)$ analyze the
simulated datasets resulting from such
observations. The purpose of this first paper is to show how these
tools can be used to evaluate the detectability of single planets
around single stars and their measurability in terms of mass and
orbital characteristics, as a function of both SIM mission
parameters and properties of the planet. In the second paper we will 
address the issues of the detectability and measurability of systems 
of planets with SIM, as well as extensive analyses of the 
instrument capability 
to determine the coplanarity (or non coplanarity) for a variety of 
orbital arrangements in multiple-planet systems. 

The first paper is organized as follows. In the second Section we
briefly describe SIM narrow-angle astrometric mode. In the third
Section we present a description of the software for the simulation
of SIM narrow-angle observations.
Details on detection and orbit determination methods are given in
the fourth Section. The most significant results obtained so far
are presented in the fifth Section, followed by summary and
conclusions.

\section{SIM Narrow-Angle Astrometric Model}

In its current design, the SIM instrument consists of three 
Michelson interferometers
(each composed of two starlight collectors, a beam combiner, and
a delay line) with nearly parallel baselines, simultaneously operated
at optical wavelengths (0.4$-$0.9 $\mu$m).
Eight starlight collectors (siderostats) are distributed
along the spacecraft structure (two provide redundancy), so that
at any time three interferometers with three different baselines
can be selected. An external metrology truss monitors the relative
positions of the three baselines. While one interferometer, the
``science'' interferometer, measures angular positions of the
celestial objects in the observing plan, two other
interferometers, the ``guide'' interferometers, observe bright
{\it guide stars} close to the target of interest, to stabilize
the orientation of the interferometric baseline. The presence of
the two guide interferometers is required for the SIM co-linear
architecture, as space-based instruments obviously do not have the
benefit (and encumberences) of a rigidly rotating stable platform
from which to operate (the Earth). Thus, to
anchor the spacecraft attitude to a celestial coordinates system,
guide interferometers lock on two bright reference stars.
During the period of time in which SIM baseline
is stabilized ($\sim 1$ hr), the science interferometer can
observe targets within a 15 degree-diameter Field of Regard (FoR).
The area of the sky on which a set of objects is observed with
the science interferometer, while
the guide interferometers determine the orientation of the science
baseline in inertial space, is called a {\it tile}. Knowledge of the
absolute attitude is actually not essential. What matters is that
changes in the baseline orientation during the observing period
are accurately monitored. The initial length and orientation of
the baseline are estimated in the post-processing of the astrometric
data.

Global astrometric missions, such as Hipparcos, DIVA, and GAIA,
are designed to perform differential angular measurements in their
sensitive directions between stars observed at about the same time
 by centroiding their diffraction-limited images.
The basic astrometric observable for SIM is the optical path-length
difference between the two arms
of the science interferometer, once it is ``locked'' on the
fringes of a target star. The optical path-length delay $d_{\star}$
can be functionally related to the spacecraft attitude and the 
three-dimensional 
instantaneous position on the sky of the target of interest by
means of the fundamental astrometric observation equation:

\begin{equation}\label{delay}
  d_{\star} = \mathbf{B}\cdot\mathbf{S}_{\star} + C
  +\sigma_{\star}
\end{equation}
where $\mathbf{B} = B\mathbf{u}_\mathrm{b}$ is the baseline vector
of length $B$, $\mathbf{S}_{\star}$ is the unit vector to the star
being observed, $C$ is a constant term representing residual
internal optical path differences, and $\sigma_{\star}$
is the single-measurement error on the target position (expressed
in picometers). All delay measurements with the science
interferometer of objects within the same tile, during the 1 hr
observing period, are made with the SIM spacecraft inertially
pointed, i.e. while the guide interferometers are used to keep both
$\mathbf{B}$ and $C$ constant (at the $\mu$as level). The external
path delay is determined by means of the introduction of an
internal movable delay line, which serves to equalize the optical
path-length of the light beams coming from the the right and left
arm of the science interferometer to the combination point in the
beam combiner. The peak of the interference pattern in the fringe
detector occurs when the internal path delay equals the external
path delay. The interferometer is sensitive only to the component
of the star position that is parallel to the baseline. Thus, each
measurement is strictly one-dimensional, in the line defined by
the intersection of the plane of the sky and the plane define by the
baseline vector and the direction to the star under investigation.
The necessary two-dimensional information required to fully
determine the position of the object is obtained by making
observations with different (possibly
orthogonal) orientations of the baseline vector $\mathbf{B}$.

SIM's basic set of star measurements happens within a tile with a
$15^\circ$ field. Adjacent tiles overlap each other to establish
relative object positions and geometric continuity. Tiles are
linked together by stars in the overlap regions, thereby covering
the entire sky with a systematic, interlaced brick-work-like
pattern of discrete pointings. Within each tile, SIM will observe
science targets, but will also observe grid stars. The purposes of
the grid stars are to provide links to a global astrometric
reference frame, and to determine the attitude of the baseline.
The astrometric reference grid, the key to SIM's wide-angle
astrometry, will comprise approximately 3000 stars uniformly
distributed on the sky, so that there will be about 10 grid stars
per tile, which will be observed together with science targets
during the 1-hr observing periods, as well as in dedicated
periodic grid observing campaigns. The astrometric grid catalog
will have an internal accuracy of 4 $\mu$as, over two orders of
magnitude better than the Hipparcos catalog. However, it is with
SIM relative (narrow-angle) astrometry that it will be possible to
achieve the highest measurement accuracy. SIM's differential mode
will be best suited for science programs where global references
are not required, but the highest possible differential
performance is needed, as is the case for the detection of
astrometric signatures of planetary companions to nearby stars.
The objective of 1 $\mu$as relative astrometry in 1 hr integration
time corresponds (see Eq.~\ref{delay}) to an accuracy on the
position of the delay line of 50 pm with a 10 m baseline. Relative
one-dimensional laser gauge metrology for high-precision measurements
and control of changes in baseline length has been demonstrated in
laboratories at a level of a few picometers on short 
time-scales~\citep{gursel93,noecker95,reasenberg95,leitch98}, 
while ongoing three-dimensional experiments at JPL, such as 
the Micro-Arcsecond Metrology Testbed~\citep{Shaklan98,kuhnert98}, 
are currently progressing towards the 50 pm goal for SIM~\citep{kuhnert00}. 
Finally, the Micro-Precision
Interferometer Testbed at JPL has already achieved the 10
nm positional stability requirements on internal optical
pathlengths~\citep{neat98}, necessary to ensure the
maintenance of fringe visibility.

While operated in narrow-angle astrometric mode, during an
observing period within a single tile,
the science interferometer carries out delay measurements of a
target star and a number of nearby reference stars, located within
a circle $\sim 1$ degree in diameter, centered on the target. The
fundamental measured quantity is then the {\it relative} delay:

\begin{equation}\label{relative}
  \Delta d_{\star,n} =
  \mathbf{B}\cdot(\mathbf{S}_{\star}-\mathbf{S}_n)+
  \sigma_{\mathrm{d}}
\end{equation}
which corresponds to the instantaneous angular distance between
the target in the observing plan and its $n$-th reference star,
projected onto the interferometer baseline, while $\sigma_{\mathrm{d}}$
is the single-measurement accuracy on each relative delay measurement.
Note that, to first order, the constant term $C$ cancels out
for the set of relative delay measurements within the 1-hr
observing period.

\section{Simulations of SIM Observations}\label{three}

The simulation and analysis code was developed to
perform studies similar to those conducted by 
Reasenberg et al.~\cite{reasenberg97}, Lattanzi et al.~\cite{latt97}, 
Sozzetti et al.~\cite{sozzetti00}, and Lattanzi et 
al.~\cite{latt00a,latt00b}. It has been specifically tailored 
to reproduce SIM observations in narrow-angle mode.

\subsection{Stellar Distribution and Model of Motion}

All our simulations use a consistent set of definitions to build a
sample of targets with given properties. We start by generating
200 science targets at uniformly-distributed random locations on
the celestial sphere. Each defines the center of a circular region
({\it domain}) of about $1^\circ$ in diameter, which is the effective
field of view for narrow-angle astrometry. Then we generate guide
and reference stars for each target. The two guide stars, which
ensure the stability of the interferometric baseline during the
course of each observation, are placed near the border of each
domain, and a predefined number of reference stars is placed
randomly within this region. For each target, reference, and guide
star, we express the five basic astrometric parameters (two positions
$\lambda$ and $\beta$, two proper motion components $\mu_\lambda$
and $\mu_\beta$, and parallax $\pi$) in ecliptic coordinates. The
parallax of the target star is predetermined by each experiment
(i.e., the star is placed at a specific distance from the Sun),
while the parallax of the reference stars is chosen randomly around 1 mas.
Proper motions are generated randomly from a normal distribution with
dispersion appropriate to each star---larger for the target, which
is assumed to be nearby, and smaller for the more distant
reference stars. The introduction of a detailed Galaxy model is a
possible future enhancement. However, detectability and
measurability of planets are essentially independent of the values
of all proper motions and of the parallax of the reference stars.
The unperturbed photo-center positions of the target and reference
stars are then computed on the basis of the five astrometric
parameters of each star. Then, we correct the photo-center position
of the target for the
gravitational perturbation induced by the presence of a planetary
mass companion. The Keplerian motion of the orbiting planet is
described via the full set of seven orbital parameters: semi-major
axis $a$, period $T$, eccentricity $e$, inclination $i$, longitude
of pericenter $\omega$, position angle of the line of nodes
$\Omega$, and epoch of pericenter passage $\tau$. 
The target's actual orbital motion around the sytem barycenter is then 
obtained by scaling the planet's semi-major axis by the ratio of masses 
and the distance.

To describe the target motion on the celestial sphere, we have
adopted a linear analytic model, in which the difference between
the position vector to the target $\mathrm{S}_{\star}$ evaluated
at time $t$ and the same quantity $\mathrm{S}^\prime_{\star}$
measured at time $t^\prime$ is expressed as a sum
of small perturbative terms to the initial location of the target
on the sky, due to proper motion, parallax, and the gravitational
perturbation induced by the orbiting planet:

\begin{equation}\label{model}
\mathrm{S}^\prime_{\star}-\mathrm{S}_{\star}={\rm d}{\bf
S}_{\star} = {\rm d}{\bf S}_{\mu}+{\rm d}{\bf S}_{\pi}+{\rm d}{\bf
S}_{K}
\end{equation}

In this model, second order effects such as relativistic
aberration and light deflection from the major solar-system bodies
are not taken into account, nor have we considered other secular 
changes in the target's position due for example to changing proper 
motion (perspective acceleration) or changing parallax. 
Realistically, such effects will have to be taken
in consideration in future experiments. However, in these studies
we assume that {\it a-priori} corrections for higher order effects
have been made to the simulated observations which we use as input
to the detection and analysis part of the code.
Also, targets and reference stars are assumed
astrometrically clean: sources of astrometric noise such as flares or
spots on the stellar surface have not been modeled throughout our
simulations, nor have we discussed the possibility of binaries
among either targets or reference stars, except for the single planet 
orbiting the target star.

Finally, we generate a set of narrow-angle observations of each
system at predefined times and baseline orientations, adding
measurement errors as discussed below.

\subsection{Error Model and Observing Scenario}\label{scenario}

Currently, SIM narrow angle error budget for each one-dimensional
visit assumes that a 1 $\mu$as accuracy on one axis will be
reached for both target and reference stars, with errors scaling
as the square root of the exposure time $\sqrt{t}$, after a
nominal 1 hr ``performance specification period'', which comprises
spacecraft slewing between adjacent wide-angle FoRs, stabilization
and acquisition of two bright guide stars in order to allow the
instrument to maintain the correct attitude during observations
within that FoR, observations of a number of grid stars, which are
used to remove instrumental thermal drifts, and sequences of
elemental fringe measurements of the target and its reference
stars, which are called {\it unit observing blocks}. Within each
unit observing block, the science interferometer executes
elemental delay measurements of the target and reference stars in
turn, with a minimum integration time per object of 30 sec, needed
for metrology stabilization. The effect of thermal drifts is such
that the maximum duration of each observing block is currently
constrained to be no greater than 5 minutes, even if a thorough
understanding of the tradeoff between longer integration times
(and increased efficiency) and possible performance degradation
with longer time-scales has not yet been achieved.

At present, knowledge of error sources aboard SIM is still
incomplete, nevertheless it is possible to summarize the error
contribution to narrow-angle astrometric measurements as composed
of two parts, a photon noise component and an instrument
systematic component: the former is mainly a measure of instrument
throughput, including for example contributions from detector
quantum efficiency, mirror reflectivity, and imperfect fringe
visibility; the latter represents the sum of known and estimated
contributions to the optical path difference (delay) which are
instrumental in origin. The system noise and photon response
contribute to the single-measurement accuracy as (see for example
the official SIM website, \url{http://sim.jpl.nasa.gov}, and
documentation therein):
\begin{equation}\label{errorbudg}
  \sigma = \sqrt{\sigma^2_{\mathrm{sist}} +
  \sigma^2_{\mathrm{phot}}\times 10^{\frac{V-V_0}{2.5}}\times (t_0/t)}
\end{equation}
where a reference photon sensitivity $\sigma_{\mathrm{phot}} = 4$
$\mu$as is reached in $t_0 = 14.1$ hr for a star of magnitude $V_0
= 20$ mag. For a system noise $\sigma_{\mathrm{sist}} = 1.718$
$\mu$as, and for a target and reference stars brighter than $V =
11$ mag, in $t = 30$ sec integration time it is possible to
achieve an astrometric accuracy of 3 $\mu$as on each object. In
principle, at the end of 10 unit observing blocks a 1 $\mu$as
precision is reached on one axis, for each object. This simplified
noise model does not account for any term with an explicit
dependence on the angle to the reference star
$\vartheta_\mathrm{R}$. This error is due to the so-called ``beam
walk'', as the delay line is slewed and the siderostat is rotated,
and is expected to grow linearly with $\vartheta_\mathrm{R}$.

In this exploratory study we have implemented a simplified
observational scenario, in which $n$ reference stars are initially
placed around each target within the $1^\circ$ domain. 
A one-dimensional observation, which we 
will call {\it Standard Visit} hereafter, 
consists of a sequence of elemental fringe
measurements by the science interferometer of a target star and
its reference stars, while the two pointing interferometers are
locked onto the bright guide stars. In reality, observations of
grid stars (4 at least) should be executed at the beginning and
end of the Standard Visit to measure the size and orientation of the
baseline with the needed accuracy. For the moment, we do not
include grid stars observations in our simulations, and assume the
orientation and length of the interferometric baseline are
perfectly known. The structure of each standard visit is then
defined as follows. We carry out $N_b$ unit observing blocks (not to 
exceed the 1-hr ``performance specification period''), each composed
of $n$ 30-sec measurements of the target and one 30-sec
measurement for each of the $n$ reference objects. Assuming a
typical 30-sec repositioning time, each unit observing block lasts
$30\times n+30\times n+2\times 30\times n$ sec = $2\times n$
minutes. We satisfy the underlying constraint of 5 minutes limit
duration of each set of measurements of target and reference
stars, due to thermal drifts, in that each pair of
target/reference star elemental fringe measurement is completed in
only 2 minutes. The Standard Visit can thus be regarded as a
sequence of $n$ {\it differential} measurements between the target
and each of the $n$ reference stars, of the form: T$_1$ $-$R$_1$,
T$_2$$-$R$_2$,$\dots$, T$_n$$-$R$_n$.

If we assume independent errors in the measured positions of target
and reference stars (in our present analysis we do not account for
possible correlations), then at the end of each one-dimensional Standard
Visit the single-measurement error $\sigma_{\mathrm{d}}$ on
each of the $n$ relative delays will be:
\begin{equation}\label{differential}
\sigma_{\mathrm{d}} = \frac{\sqrt{\sigma^2_{\star,n} +
\sigma^2_{R,n}}}{\sqrt{N_{b}}}
\end{equation}
with $\sigma_{\star,n}$ and $\sigma_{R,n}$ the elemental fringe
measurement errors on each object defined by Eq.~\ref{errorbudg}.
Assuming the target and its $n$ reference stars are bright ($V
\leq 11$), then for each 30-sec elemental fringe measurement we
obtain $\sigma_{\star,n} = \sigma_{R,n} = 3$ $\mu$as. Finally, we
set $N_b = 4$, thus each of the $n$ differential measurements at
the end of a Standard Visit (from Eq.~\ref{differential}) has an error
$\sigma_{\mathrm{d}} \simeq 2$ $\mu$as. We will refer to the {\it relative} 
delay accuracy $\sigma_\mathrm{d}$ as the Standard Visit accuracy 
throughout the rest of this paper. 

Standard Visits of targets and reference stars can be made at
arbitrary times and arbitrary spacecraft orientations, throughout
the 5-year nominal mission lifetime--we neglect at this point
possible visibility, orientation, and planning constraints that
might restrict the observing sequence
\footnote{The SIM spacecraft is currently supposed to be operated in an 
Earth-trailing orbit, thus avoiding Earth's occultations that would occur 
in a low-Earth orbit. Only a Sun exclusion angle (presently set to 
43$^\circ$) 
will introduce restrictions on the observing schedule. The Sun exclusion 
zone will primarily affect the periodic grid observing campaigns, 
for which different scenarios have been evaluated. 
The possible impact of the Sun exclusion zone on actual observing strategies
for science targets and the limitations it may impose on the detectability 
of planets with a given period will ultimately be assessed only after a 
definite choice for the grid observing campaigns has been made}. 
Each visit is strictly
composed of one-dimensional measurements, then additional
observations with orthogonal orientations of the baseline are
required. Thus, we define a {\it full two-dimensional astrometric
observation} as the sum of two (one-dimensional) Standard Visits
executed with baselines at approximately right angles with each
other. The time separation between pairs of Standard Visits is
chosen randomly between 0 and 5 days.

\section{Data Analysis Methods}\label{four}

This section outlines the data analysis procedure we have
developed and implemented in the analysis part of the code to
assess SIM's ability to discover and measure planets.
For each case, planet detectability is measured via a standard
$\chi^2$ test of the null hypothesis that there is no planet. If
deviations in the observation residuals exceed a predefined
significance threshold, then the planet is considered detected,
and its orbit is determined by a full non-linear least-squares fit
to the observed relative delay measurements. The fit also
determines the relative positions, proper motions, and parallaxes
of target and reference stars, which are of course not known {\it
a priori}. We derive errors in the measured orbital parameters
empirically, by comparison of the values determined from the fit
with the true (input) values.

\subsection{Planet Detection}

To determine planet detectability, we conduct a null
test: we {\it assume} that the target is a single star with no
companions, and
determine whether (and with what confidence) this assumption can
be proven false. This process has two steps: first, determine the
astrometric parameters of target and reference stars under the
no-planet assumption; second, compute the discrepancy between the
best no-planet solution and the actual measurements. A detectable
planet will cause a statistically significant discrepancy.

Since all measurements are differential, positions, proper
motions, and parallaxes of target and reference stars are not
fully constrained by the time series of measured delays.
Specifically, the least squares problem has a rank deficiency of
5, corresponding to the five parameters that cannot be determined
by the observations: two position zero points, two proper motions
zero points, and a parallax zero point (in the narrow-angle
approximation). A convenient solution is to choose a so-called
{\it base object} among the set of reference stars, and to hold
its astrometric parameters fixed at their starting values
throughout the fitting procedure. Then, astrometric parameters of
all other objects are measured relatively to those for the base
object. This approximation may break down for sufficiently large
parallaxes and field angles, since parallaxes at different places
in the sky are not exactly additive. However, as long as an
approximate {\it absolute} parallax can be determined from the
grid solution, we expect the errors introduced by this
approximation to be negligible.

If the position of the target star is perturbed by a planet, the
resulting no-planet solution will have post-fit residuals large compared 
with observational noise. These observation residuals will contain 
systematic deviations due to the companion. A simple way of assessing 
planet detectability is to apply a standard
$\chi^2$ test to the residuals of the single-star fit, assuming
that the uncertainties of the individual measurements are known.
We compute the probability $P$($\chi^2\geq\chi^2_o$) of obtaining
$\chi^2$ greater or equal to that observed ($\chi^2_o$), assuming
the single-star model to be correct: if such probability is greater
than a given threshold, the {\it acceptance level}, then the observations 
are consistent with the assumption of no orbital motion; 
if $P$ drops below the acceptance level, then the single-star model is
rejected, and the planet is considered detected at the
corresponding confidence level.  We set an acceptance level
of 5\% -- and a corresponding confidence level of 95\% -- throughout
our simulations. By definition, 
this method only measures significant deviations from an {\it a
priori} model, and does not provide elements for a direct
identification of the nature of the observation residuals. To this
end, residuals are usually inspected by means of standard
periodogram analyses. In our simplified scenario, the star+planet model 
fits the data well and therefore the planet can be considered detected 
(later on in Section~\ref{errors} we provide quantitative examples). 
However, in more 
realistic circumstances, the relative quality of the single-star and 
star+planet fits should be compared, and a statistical test applied to 
determine the true significance of the detection. 
Furthermore, the level of detection threshold in the presence of 
real-world disturbances can be a source of concern. Indeed, the 95\% 
confidence level which 
was adopted in our tests may well be too liberal to indicate detection in 
the presence of realistic astrophysical and instrumental noise. A realistic 
threshold for truly reliable detection can only be established once a more 
complete model of instrumental and astrophysical effects is available. 
In the context of this exploratory study, we choose a confidence level 
of 95\% because it leads to manageable numerical experiments. 
We will discuss possible
enhancements in planet detection methods in a future work.

\subsection{Orbit Determination}

Once a planet is detected, the goal is to determine its orbital
characteristics and mass. We therefore expand our fitting
procedure to allow for the presence of a planet around the target
star. The photo-center motion of the target includes the
gravitational perturbations due to a planet, whose seven Keplerian
elements---which fully describe its orbital motion---are added to
the list of unknowns to be determined. In evaluating the observation
residuals, we employ an analytic model in which the
computed relative delay between the target and its $n$-th
reference star is in the form:

\begin{equation}\label{omenc}
  \Delta d_{\star,n} =
  \mathbf{B}\cdot(\mathbf{S}_{\star}(\lambda_\star,\,\beta_\star,\,
  \mu_{\lambda,\star},\,\mu_{\beta,\star},\,\pi_\star,\,
  X_1,\,X_2,\,X_3,\,X_4,\,e,\,T,\,\tau)-\mathbf{S}_n(\lambda_n,\,\beta_n,\,
  \mu_{\lambda,n},\,\mu_{\beta,n},\,\pi_n))
\end{equation}

For convenience, we use the Thiele-Innes representation of the
orbital parameters, in which $a$, $i$, $\omega$, and $\Omega$ are
combined to form the four Thiele-Innes elements $X_i$,
$i=1,\dots,4$ (see for example Green~\cite{green85}). The Thiele-Innes
representation is better behaved for fitting purposes, as it
reduces to three the number of non-linear parameters in the
observation equations, thus improving the convergence speed of the
iterative least squares algorithm. At the end of the
simulation, the classic orbital parameters are recomputed from the
Thiele-Innes elements.

Note that the mass of the planet is $not$ determined directly from the
orbital parameters. In fact, the planet mass can only be determined if
an independent estimate of the mass of the parent star is available,
for example, from its spectral and luminosity class. The mass of the
orbiting planet would then be computed via the mass function formula:
\begin{equation}
\frac{M_\mathrm{p}^3}{(M_\star + M_\mathrm{p})^2} =
\frac{a_\star^3}{\pi^3}\frac{1}{T^2} \label{massfunct}
\end{equation}
where $M_\mathrm{p}$ and $M_\star$ are the planetary and stellar
mass in solar-mass units, $T$ the orbital period in years, $\pi$
the parallax and $a_\star$ the semi-major axis of the orbit of the
central star around the barycenter, both expressed in arcsec. The
fit to the observed relative delays determines $a_\star$, $\pi$,
and $T$ directly, while $M_\mathrm{p}$ can be computed if
$M_\star$ is known or estimated. In fact, under the reasonable
assumption $M_\mathrm{p}\ll M_\star$, we can then derive the
planet mass via the following approximate formula:
\begin{equation}
M_\mathrm{p} \simeq \left(\frac{a_\star^3}{\pi^3}
\frac{M_\star^2}{T^2}\right)^{1/3} \label{massfunc}
\end{equation}

\subsubsection{Initial parameters}\label{guesses}

We have implemented an iterative method for the solution of the
highly non-linear system of equations of condition that
utilizes the Levenberg-Marquardt algorithm~\citep{press92}.
Initial guesses for the unknown parameters are needed to start the
fitting process. The solution at step $k$ is updated with respect
to the solution at step $k-1$, then the variable $\chi^2$ (i.e.,
the sum of the square of the observation residuals) is checked for
convergence of the solution: if $\Delta\chi^2 =
\chi^2_{k-1}-\chi^2_k \leq 0.01$, iterations are stopped. A
natural choice of the condition for stopping iterations consists
of requiring that $\chi^2$ {\it decreases} by a negligible amount,
absolute or fractional (if the opposite behavior is observed,
then further iterations are necessary for stabilization of the
solution). As a matter of fact, because the minimum of $\chi^2$ is at
best only a statistical estimate of the fitted parameters, a
variation in the parameters that changes $\chi^2$ by a quantity
$\ll 1$ is generally not statistically meaningful, as with this
method the parameters, once they approach the configuration which
minimizes the fitted function, tend to wander around in the
vicinity of the minimum, in a flat valley of complicated
topology.

In any iterative, non-linear fitting procedure, the choice of the
initial guess for the unknown parameters can have a significant
impact on the convergence of the method and the quality of the
solution, especially because of the possibility of false (local)
minima of the $\chi^2$ function. Unlike Konacki et al.~\cite{konacki02}, 
since we are focusing
primarily on the performance that SIM can ultimately achieve in
detecting and measuring planets, the orbital fitting results we
discuss in the next section used {\it good} guesses, which means
they were obtained with initial guesses differing from the
``known'' values of each parameter by only a moderate amount of noise.
The deviations of the resulting fitted parameters from their true
values should then be a useful measure of the accuracy in orbit
reconstruction that can ultimately be achieved with the assumed
measurement errors. Note also that our detection estimates are
intrinsically independent of the choice of initial parameters,
since the detection least squares problem is linear in all fitted
parameters (in the narrow-angle approximation), and therefore its
solution is independent of the initial parameters.

Nevertheless, a proper assessment of the effectiveness of any overall
search and measurement strategy requires a more realistic
approach. Initial parameters must be determined solely on the
basis of the actual measurements, without any {\it a priori}
knowledge of the system, and double-blind tests must be conducted
to verify the global performance of the search and analysis
method. Work is in progress, and will be presented in the future,
on refined models for global search
and optimization strategies of starting guesses for orbital
parameters, where we take into account the results from standard
periodogram analysis as well as detailed Fourier analyses of the
astrometric signal.

\section{Results}

The observation-modeling and analysis software we have developed
have allowed us to provide a first quantitative estimation of the
ability of SIM to detect and characterize the orbits of planets
around nearby stars. Preliminary findings were shown by
Casertano \& Sozzetti~\cite{caser99}. The more general and
complete set of results we present in this Section has been
obtained using a number of {\it a priori} assumptions (see
Sections~\ref{three} and~\ref{four}) on the instrument
(perfect knowledge of the error model and of the
satellite attitude), on the systems to be investigated
(astrometrically clean targets and reference stars, no stellar
companions, single planets), and on the data analysis procedures
(avoiding studies of periodicities and double-blind tests, and 
utilizing good guesses for the values of the orbital parameters 
necessary to initialize the least squares solution). 
The main focus of this work is on the goal of determining 
SIM's {\it ultimate} ability to detect and measure single planets around 
single, normal, nearby stars, and the above assumptions constitute 
the most efficient way to achieve it. 
We expect that some of these assumptions will have a non-negligible 
impact on the actual planet-finding capabilities of SIM, which should 
be revisited when a more realistic description of the satellite and 
its operations becomes available. 

In this Section, we present and discuss our most significant
results, as follows. First, we compute detection probabilities and
estimate the accuracy achievable in measuring the orbital elements
and mass of a planet as a function of its characteristics and of
SIM single-measurement precision, utilizing a template observing
strategy based on the simplified
observational scenario sketched in Section~\ref{three}. Next, we
study the performance of different observing
strategies, varying the number and time-spacing of observations,
and the number of available reference stars around the target.
Finally, we discuss the merit of more flexible observing scenarios 
to be applied to both bright ($V\leq 11$) and faint targets, and 
utilize them to identify
the boundaries of the discovery space of SIM for detection of 
terrestrial planets around a sample of the nearest solar-type stars 
and M dwarfs.

\subsection{Detection Probabilities}\label{detect}

Astrometric observations of a star contain its reflex motion due
to orbiting bodies, including planets. The apparent magnitude of
this perturbation, the orbital motion of the star around the
center of mass of the system, is the so-called {\it astrometric
signature}:
\begin{equation}\label{signature}
  \alpha =
  \frac{M_\mathrm{p}}{M_\star}\frac{a_\mathrm{p}}{D},
\end{equation}
where $M_\mathrm{p}$ and $M_\star$ are, respectively, the mass of
the planet and the central star, $a_\mathrm{p}$ is the semi-major
axis of the planet's orbit, and $D$ the distance of the system
from the observer. If $M_\mathrm{p}$ and $M_\star$ are given 
in solar mass units, $a_\mathrm{p}$ in AU, and $D$ in parsec, then
$\alpha$ is in arcsec.

In principle, detection probability will depend upon $a)$
measurement errors due to correlated and uncorrelated instrumental
and astrophysical noise sources, $b)$ mission parameters, and $c)$
distance and properties of the observed star-planet systems. We
have discussed in section~\ref{three} and~\ref{four} the basic
assumptions we have made while taking into account the
contributions to points $a)$ and $b)$. As for point $c)$, we will
express detection probabilities as function of the orbital period
$T$ of the planet, the distance $D$ from the observer, and the
astrometric signature $\alpha$. In particular, our simulations are
based on an assumed Standard Visit accuracy $\sigma_\mathrm{d} = 2$
$\mu$as for each relative delay measurement, which applies to
bright targets and reference stars ($V \leq 11$ mag), in the
context of the current best-estimate error budget for SIM
narrow-angle astrometric observations, and for a structure of the
Standard Visit as discussed in Section~\ref{scenario}. In order to
apply our results to a different single-measurement accuracy, we
note that the detection probability depends on $\alpha$ and
$\sigma_\mathrm{d}$ only through their ratio, which we call the
``scaled signal'':
\begin{equation}\label{signal}
  S = \alpha/\sigma_\mathrm{d}
\end{equation}
Similar scaling applies to the precision with which the orbital
parameters can be determined. Thus, the results we present here
can be easily rescaled to different measurement errors. For given
scaled signal, the detection probability depends, of course, on
all other orbital parameters, and especially on the period $T$ of
the orbit.

First of all, we ran a set of simulations without planets,
generating spheres of 200 uniformly distributed targets, and observed
them with a template observing strategy, in which a specified
number of reference objects is chosen for each target, together with
a fixed number and time-spacing of pairs of one-dimensional orthogonal
observations, during the nominal mission duration. 
For the purpose of our analysis, we have placed $N_r = 3$ 
astrometrically clean reference
stars within a $1^\circ$ domain centered around each target, and
adopted sequences of $N_o = 24$ two-dimensional (orthogonal) narrow-angle
observations (as defined in Section~\ref{scenario}) equally spaced over
the 5-yr mission duration, with a time interval between pairs of successive
observations fixed to 0.2 years, and assuming that a full observation
is completed within 5 days. We thus have a total of $N_{m} = 144$ 
relative delay measurements, and a total of $N_{p} = 20$ unknown parameters 
to solve for in the set of observation equations,
i.e. five astrometric parameters for the target and each of the components of
its local reference frame. Within this template observational scenario, 
a Standard Visit lasts about half an hour, 
well within the recommended 1-hr limit. 
The observing sequence described above was utilized to obtain all
simulation results discussed in this and the following section.
We have verified the correct behavior of the $\chi^2$ test and the
choice of the confidence level, as described in the previous Section.
As expected, the number of {\it false detections} was $\simeq 5\%$.
Then, we have used the $\chi^2$ test to analyze 320\,000 1-$M_\odot$
stars uniformly distributed on the sky, orbited by
single planets producing astrometric signatures in the range
$1\leq\alpha\leq
40$ $\mu$as, with periods in the range $0.5\leq T\leq 20$ yr. We
averaged over the remaining orbital elements, distributed randomly
in the ranges: $0^\circ\leq i\leq 90^\circ$, $0\leq e\leq 1$,
$0\leq\Omega\leq \pi$, $0\leq\omega\leq 2\pi$, $0\leq\tau\leq T$.

Figure~\ref{fig1} summarizes the main characteristics of the
planet detection probability. As discussed above, we consider a
planet detected when the null-test, based on the $\chi^2$ of the
astrometric solution for target and reference stars assuming that
there is no planet, fails at the 95\% confidence level. The curves
in Figure~\ref{fig1} represent equal probability contours for the
detection; thus, for example the curve marked 95\% indicates the
locus of the period $T$ and scaled signal
$S$ for which 95\% of the planets generated
in our simulations fail the null-test, at the 95\% confidence
level. The minimum astrometric signature required to achieve secure
detection must be $\alpha_\mathrm{min}\sim 2.2$ times the
Standard Visit accuracy $\sigma_\mathrm{d}$
(we recall that accuracy refers to independent measurements of
relative delays between the target and each of its reference
stars), for periods between 0.5 and 5 years. Due to the
increasingly worse orbital sampling, the required signal rises
sharply for periods longer than the mission length, especially for
high detection probabilities. However, a Jupiter-Sun system, with
a period of 11.8 years, can still be detected 50\% of the time if
the astrometric signature is about $4\,\sigma_\mathrm{d}$,
or to a distance of 500 pc.

The dashed lines indicate the signature produced by systems composed
of a 1-$M_\odot$ primary and a 20-$M_\oplus$ planet placed at the
distance shown in the legend, as a function of orbital period (in
parentheses, the equivalent distance at which
a system composed of a solar-mass star and Jupiter-mass planet would
produce the same signature). These curves are derived by substituting
Kepler's third law in the defining expression for $\alpha$
(Eq.~\ref{signature}), in the limit for $M_p\ll M_\star$:

\begin{equation}
\alpha\simeq \frac{M_\mathrm{p}}{M_\star^{2/3}}\frac{T^{2/3}}{D}
\label{kepler}
\end{equation}
Thus, for example, a Jupiter-mass planet at 300 pc (long-dashed lines),
or equivalently a Neptune-class planet placed at
20 pc, can be detected with 95\% probability around a 1-$M_\odot$
star if its orbital period is roughly between 1.5
and 8.5 years---the range over which the long-dashed line
lies above the 95\% contour.

A different representation of the same results is shown in
Figure~\ref{fig2}, where the isoprobability contours are drawn as
function of the period and distance, for a Jupiter-mass planet
around a solar-mass star. The maximum detection distance peaks for
$T\simeq 4$ years; shorter periods correspond to smaller orbital 
amplitudes, and thus smaller values of 
$S$, while planets with longer periods
suffer from increasingly incomplete orbital sampling.

In Figure~\ref{fig3} we show the behavior of the
detection probability as function of $S$,
for different orbital periods. As already emphasized by Reasenberg
et al.~\cite{reasenberg97} and Lattanzi et al.~\cite{latt00a}, who 
studied the planet-finding capabilities of the proposed NASA mission POINTS
and of the recently approved ESA Cornerstone Mission GAIA, planets 
with periods longer than the mission lifetime require a much stronger 
signal in order to be detected with high confidence. 
Finally, this plot highlights how, within the range of
favorable periods ($0.5\leq T\leq 5$ yr), detectability in the
case of $T = 1$ year (dashed-dotted-dotted line) is slightly
affected by the coupling between orbital and parallactic motion.

\subsection{Estimation of Planet Mass and Orbital Elements}\label{estimate}

Just as the detection probability (for a given choice of the observing
strategy), the ability in measuring accurately the set of orbital
parameters for a detected planetary
system is closely related to the period and the astrometric signature
induced by the unseen planet on the observed parent star.

First, we have utilized our simplified but realistic observational
scenario to provide an estimate of the rms errors expected on the
determination of orbital parameters and masses in a handful of
significant cases, together with an evaluation of the quality of
fits that may be obtained. Next, we have determined the
boundaries, in the $\alpha-T$ plane, of SIM ability to accurately
determine the orbital geometry and mass of a single planet
orbiting a single normal nearby star, by evaluating the minimum
astrometric signature required for measurements of a given orbital
parameter or the mass of the planet good to a given accuracy
level. Finally, we have considered how the presently
known candidate planets would fall within the limits of SIM's
detection and orbit reconstruction capabilities. The extra-solar 
planets found so far constitute a natural laboratory for conducting 
physical studies of planetary systems with SIM. In fact, astrometry 
measures two projections of an orbit, as opposed to the intrinsically 
one-dimensional radial-velocity measurements, and thus can 
determine the entire set of 7 orbital parameters. To this end, 
SIM's highly accurate measurements will be instrumental in
breaking the inclination degeneracy intrinsic to radial velocity
observations, and this will allow in turn a direct estimate of a
planet's true mass. As discussed in detail below, M$_\mathrm{p}$ and $i$ 
are two key parameters to be determined for a proper understanding 
of the nature and diversity of sub-stellar companions.

\subsubsection{Empirical Errors and Quality of Fits}\label{errors}

Our simulations cover four particularly significant 
cases: $1)$ a Jupiter-mass
planet in orbit around a solar-mass star with $T=1$ year at
$D=100$ pc, to quantify the effect of the coupling between parallactic
factor and orbital period when attempting the orbit reconstruction; 
$2)$ the same system but with an orbital
period $T=5$ years at $D=200$ pc, in an almost ideal configuration
where the orbital period equals the mission duration;
$3)$ a short-period ($T$ = 15 days) giant planet
($M$ = 4 $M_\mathrm{J}$) around a 1-$M_\odot$ at $D$ = 30 pc, to test
SIM's ability to cope with poorly sampled motion;
$4)$ a ``true'' Jupiter-Sun system, with a period of
12 years---over twice the mission duration---at $D$ = 100 pc, to
stretch SIM's ability to solve long-period orbits.
These four systems have scaled signals $S$ = 5, 7, 8, and 25, respectively.

In all cases, utilizing the template observing scenario described in 
Sections~\ref{scenario} and~\ref{detect}, 
we have simulated SIM observations of 200 systems
uniformly distributed on the sphere, averaging over the
remaining orbital parameters ($i$, $e$, $\tau$, $\omega$, and $\Omega$). 
We have then modeled a single-star fit on the simulated observations,
and in all cases excessive residuals indicated (with a 95\% confidence
level) the presence of a companion. Finally, we fitted the measured
perturbation of the unseen secondary
with a full Keplerian orbit. The deviation of the fitted
parameters from their true values gives an indication of the
accuracy of the measurements. The results for three of the four cases
mentioned above are presented in Figure~\ref{fig4}. For each parameter,
the rms deviation of measured from true value is given in its respective
panel.

The general indication is that a scaled signal
$S\simeq 5$ (minimum astrometric signature 5 times larger than the
single-measurement precision) is sufficient to measure the
parameters of the system with an rms error of about 20-30\%, as
long as at least one entire orbit is observed during the mission
lifetime. If $T > 5$ yr, a stronger signal is required, and the error
distribution can have an enhanced tail, particularly in the case of
the semi-major axis $a$, because of systems which, due
to eccentricity, orientation, and phase, have an unfavorable
orbital sampling during the time spanned by the observations.
Due to the good orbital sampling, 
the 1-yr period is recovered very accurately, and the coupling
between parallactic and orbital motion (highlighted by the broader
distribution of rms errors on $\pi$) turns out to have a less critical 
impact than one might have anticipated.

Simulations of a giant planet in a very fast orbit
around a 1-$M_\odot$ star at $D = 30$ pc are not shown in the figure,
as no reliable orbit can be obtained for this system. The planet can
be clearly detected in the observation residuals, but the 0.2-yr
sampling period is larger than the very short orbital period
($T\simeq 15$ days), and aliasing
effects cause the orbital fit to be poorly constrained. 
As we discuss later on in Section~\ref{strategy}, for detectable 
planets in very fast orbits reliable orbital fits will likely 
require the adoption of {\it ad hoc} strategies in the
distribution of the observations, to ensure proper coverage of
the short periodicity of the signal.

In Figures~\ref{fig5} and~\ref{fig6} we present graphical
illustrations of the quality of the fits that can be obtained for
star-planet systems, in the most favorable case among the
four discussed above. Specifically, the two Figures show
information for the system composed of a 1-$M_\odot$ star and
a 1-$M_\mathrm{J}$ planet placed at $D = 200$ pc, with the 
planet in a 5-yr period, eccentric ($e = 0.6$) orbit, and with 
the orbital plane inclined of an angle $i = 45^\circ$ with respect 
to the line of sight. The panels in
Figure~\ref{fig5} show the apparent motion in the plane of the
sky, and the residuals of the single-star fit. The apparent motion
is dominated by the proper motion and parallax of the primary
(along both the X- and Y-axis), and the motion due to the planet
cannot be discerned in these full-scale plots. The observation
residuals after proper motion and parallax have been subtracted
still highlight significant scatter, up to over 5 times the
Standard Visit accuracy ($\sigma_\mathrm{d}$ = 2 $\mu$as). The panels in
Figure~\ref{fig6} show the residual motion, measured and fitted,
after subtraction of the best-fit proper motion and parallax
terms, thus displaying more clearly the effect of the planet. After a
fully Keplerian orbit is modeled on the observations, the residuals
exhibit a scatter no larger than $\sim 2\,\sigma_\mathrm{d}$,
consistent with the absence of further companions, at the level 
of the single-measurement error. In
particular, in the lower right panel we can appreciate a
slight offset (in both X- and Y-axis) of at most 2-3 $\mu$as between the
true and the post-fit reconstructed orbital motion, in agreement with
the residual scatter observed in the post-fit residuals, and to be
compared with the analog rms error on the semi-major axis shown
in the upper right panel of Figure~\ref{fig4}.

\subsubsection{Accurate Orbit Reconstruction and
Measurement of Known Planets}\label{known}

To provide an overall estimate of SIM's ultimate capability in
measuring single planets around single, nearby solar-type stars,
we follow the same logic adopted in Section~\ref{detect}, in which we
discussed isoprobability curves for 95\% confidence of detection.
A minimum astrometric signature 
$\alpha_\mathrm{min}\sim 2.2\,\sigma_\mathrm{d}$ was
required for secure detection, for periods shorter than the
mission lifetime, with a sharp rise in the required signal for $T
> 5$ years. Similarly, it is possible to define the minimum
astrometric signature required for measurements of a given orbital
parameter or the mass of the planet good to a given accuracy
level. We ran simulations adopting the same observational scenario
described in Section~\ref{detect}, but letting astrometric signatures
vary in the range $1\leq\alpha\leq 1000$ $\mu$as. We have then
obtained the ``probability of convergence'' for the orbital parameters and
mass of the planet, i.e. the percentage of the values for each parameter
that in a simulation, after modeling the observations
with a full Keplerian orbit, falls within a given fraction of the
true value.The criterion for stopping
the iterative fitting procedure is the one described in
Section~\ref{guesses}. Similarly to Section~\ref{detect}, we assume a given 
orbital element is confidently estimated with, say, 30\% accuracy if 
the relative convergence probability within the same fraction of the 
true value is $\geq 95\%$. We find that, for example, for periods 
shorter than the mission length, mass measurements accurate to 10\% 
require $\alpha \simeq 10\,\alpha_\mathrm{min}$, and to measure the
inclination at the same level of accuracy we need a stronger
signal, $\alpha \simeq 15\,\alpha_\mathrm{min}$. The other orbital 
elements ($e$, $\tau$, $\omega$, and $\Omega$) are accurately determined 
when the astrometric signature $\alpha\simeq 5-15$ $\alpha_\mathrm{min}$, 
while depending on how many orbits are fully sampled, the period $T$ can 
be estimated with higher accuracy at smaller values of $\alpha$. 

The limits in SIM's ability to detect and measure planets are
better understood in terms of: $1)$ the impact of its measurements
on future planet discoveries, and $2)$ the wealth of information
its high-precision astrometric observations will provide
for a complete classification of
planetary systems. Figure~\ref{fig7} shows how, in the
plane $\alpha-T$, the set of presently known candidate
planets falls within
the boundaries for secure detection and accurate orbit and mass 
determination with SIM discussed in Section~\ref{detect} and above. 
We focus in particular on the accurate measurement of $M_\mathrm{p}$ 
and $i$, as  high-precision 
astrometry contributes decisively to the complete 
determination of the true orbit of a planet by removing the 
inclination degeneracy intrinsic to radial velocity measurements, thus 
providing a direct estimate of its true mass. 
In Figure~\ref{fig7}, the known planet candidates are plotted
for the worst case when the orbit is viewed edge-on. The
`worst case' means that the true mass and true astrometric
signature equal their radial-velocity minimum values. The real 
performance on most planets will improve, as depending on the actual 
inclination angles, the true masses and 
astrometric signatures will be stronger than plotted. For example,
if the currently unknown inclination angle of HD 177830 is found
to be $45^\circ$, instead of $90^\circ$ as assumed for the plot,
the mass of the planet is 1.4 times larger than the minimum mass
determined by spectroscopy (1.8 M$_\mathrm{J}$ instead of 1.28
M$_\mathrm{J}$). The expected uncertainty of our mass
determination is an absolute mass which we express as a
percentage--22\% in this case--of the minimum mass. Thus, in this
example, while the uncertainty in the true mass remains
essentially unchanged at 0.28 M$_\mathrm{J}$, the fractional error
improves to 16\% of the planet mass, instead of 22\%.

As a general result, within the limits of the simplified observational
scenario sketched in Sections~\ref{scenario} and~\ref{detect},
we find that $\sim 75\%$ of the present-day known planets
would be detected by SIM, and $\sim 50\%$
of them would have mass and inclination of the orbital plane
measured to 10\% accuracy or better. Independent and
accurate estimates of the true masses {\it and} inclination angles 
of the planet candidates
will provide important information that will help move in the 
direction of a thorough understanding of the nature of such sub-stellar 
companions. In fact, today there 
is still a lack of common consensus on the definition criteria of planets 
and brown dwarfs, and mass alone may not be sufficient in order to 
establish whether a low mass companion to a star is a planet, unless 
its mass is close to that of the Earth~\citep{black97}. 
For example, some of the present planet candidates 
have minimum masses close to the (somewhat artificial) 13 $M_\mathrm{J}$
dividing line between the two classes of objects set by the no-deuterium 
burning argument, and might end up belonging to 
the latter class (due to the actual values of the inclination 
angles). On the other hand, depending on whether planets and brown 
dwarfs will turn out to be formed by different or similar mechanisms, 
the 13 $M_\mathrm{J}$ cut-off itself might constitute a poor basis 
for classification,  Indeed, we might discover 
that planets and brown dwarfs share a common mass-range. For a
complete comprehension of the nature and diversity of sub-stellar
companions, other genesis indicators will have to be evaluated,
such as orbit shape and alignment of orbits in multiple systems, 
or composition and thermal structure of the atmospheres.

A closer look at Figure~\ref{fig7} shows that,
for example, the two outer planets ($\upsilon$ And c and
$\upsilon$ And d) in the three-planet system around the star
$\upsilon$ Andromed\ae\,~\citep{butler99} would both have 
mass and inclination of the respective orbital planes measured
with high accuracy, while the signal due to the presence of the
innermost planet is not even detectable. As for the other 
candidate planetary systems, in all cases at least one of 
the planets is measured with
high accuracy (10\% or better). Then, the question arises
naturally. Could SIM establish, and with which accuracy, whether
planets in a planetary system lie on coplanar orbits, or not? 
The unexpected orbital arrangement of the planets
orbiting $\upsilon$ Andromed\ae\, has for example triggered an
on-going debate on the system architecture, orbital evolution and
long-term stability~\citep{laughlin99,rivera00,stepinsky00,barnes00}. 
Then, determining
the actual planet masses {\it and} three-dimensional geometry of the
system becomes crucial in order for theory to derive sensible
conclusions on such issues. In paper II, we will
present results from extensive simulations of SIM observations of
extra-solar multiple-planet systems, in order to quantify SIM's 
capability to discover and measure systems of planets,
as well as detailed studies of SIM ability to determine 
coplanarity of multiple-planet orbits. Following similar, recent 
analyses~\citep{sozzetti01}, we will show how, 
by accurately measuring inclination angles and
lines of nodes in multiple-planet systems, SIM will provide
important observational data which will help learn about the
dynamics of evolution of planetary systems, as well as contribute
to the understanding of the intrinsic nature of sub-stellar
companions.

\subsection{Observing Strategy Analysis}\label{strategy}

The results presented in Section~\ref{detect} and~\ref{estimate}
constitute a first step towards a proper assessment of SIM
detection sensitivity and ability to accurately determine the
full three-dimensional orbital geometry and the mass of single
planets orbiting single nearby solar-type stars. Our findings have
been expressed as function of: $a)$ the most relevant physical and
dynamical parameters of the observed systems (mass, distance,
orbital period), and $b)$ the most basic instrument
characteristics. In particular, we utilized a simplified, but
realistic, observational scenario (see Section~\ref{scenario}
and~\ref{detect}), in which we kept {\it fixed} the number of
reference stars and the number and time-spacing of full
two-dimensional observations during the mission duration. These
results apply to bright targets and reference stars ($V\leq 11$),
in the context of a simple model of SIM elemental fringe 
measurement error
(see Section~\ref{scenario}), expressed in terms of the visual
magnitude and the integration time on the measured object.

In reality, the overall scientific performance of SIM operated
in narrow-angle mode, both in terms of $1)$ its potential for
unprecedented discoveries, and $2)$ its capability to
accurately measure crucial
parameters (masses, inclination of orbital planes) for a better
understanding of planet formation and evolution processes, will be
optimized only by implementing flexible observing strategies,
sometimes specifically tailored to a single target, to
maximize the ratio of the number and intrinsic scientific interest
of the objects in the target list to the fraction of the total
observing time utilized.

In this Section we analyze the performances of different observing
strategies as function of the number of reference stars per
target, the total number of full (two-dimensional) observations,
the time-spacing of epochs at which Standard Visits are executed,
and the structure of the Standard Visit itself, to account for
both bright and faint targets. This will allow us to identify a
more flexible observational scenario, which we will use in
Section~\ref{space} to delineate the borders of SIM discovery
space for detection of Earth-class planets around the
closest stars.

\subsubsection{Number of observations and reference
stars}\label{refer}

The robustness of the local reference frame relative to which a
target position is measured crucially depends on the availability
of a sufficient number of clean reference stars, for which 
all possible astrophysical astrometric noise sources (such as binarity and
surface activity) are either well certified, or negligible at the 
$\mu$as level. It is conceivable that astrometrically clean reference 
objects will be carefully selected for targets in narrow-angle mode 
before SIM's launch by accurate preparatory ground based
spectroscopic work; however, it is possible that one or more of
the selected reference stars for a given science object might turn
out to be not astrometrically stable at the level required for
very precise local astrometry. In some cases such ill-behaved
objects might have to be removed from the local reference frame.
In the event of anomalies discovered in individual reference
stars, it would be beneficial if their number (wherever it would
be feasible) was sufficiently high right from the start of SIM
observations, to increase the robustness of the local frame of
reference. This in turn means more unknowns to be solved for (five
more for each reference object), so in principle more observations
are needed to guarantee the same performance in terms of
detectability and measurability of planets.

To verify which scaling applies to detection sensitivity and
accuracy in orbit reconstruction and mass determination in the
case of a change $a)$ in the number of reference stars per target
$N_{r}$, or $b)$ in the number of full observations $N_{o}$
during the 5 yr mission, we have simulated uniform spheres of 200
Jupiter-Sun systems, with orbital period $T = 5$ yr, near the peak
in the detection probability curve in Figure~\ref{fig2}, and
averaged over the remaining orbital parameters.

For a Standard Visit accuracy $\sigma_\mathrm{d} = 2$ $\mu$as on
each relative delay, Figure~\ref{fig8} shows the maximum distance
at which, for the system described above, detection probability $P
\geq 95\%$, as a function of the number of two-dimensional
equally-spaced observations, and for 3 and 6 astrometrically clean
reference stars, respectively. As expected for a situation in
which single-measurement errors are independent, we verify a
simple scaling in detection sensitivity according to
$\sqrt{N_{o}}$. A similar scaling ($\sim\sqrt{N_{r}}$) 
holds for a variation in the number of reference stars. As
a matter of fact, more reference stars means more unknowns (5 additional 
astrometric parameters per reference object), but also more 
differential delay measurements per Standard Visit (as many as the 
number of reference stars), so that for example doubling $N_r$ or $N_o$ 
in practice produces similar effects in terms of detectability 
thresholds. The same
scaling applies to the computed errors in measured physical and
dynamical quantities of the system, i.e planet mass and orbital elements.

\subsubsection{Timing of observations}\label{timing}

The choice of the distribution of observations is crucial in many
respects: $a)$ in terms of the potential for discovery of new
planets, it would be desirable to achieve the best sensitivity 
over a wide range of periods, from a few days to the 5-yr SIM-mission 
duration; $b)$ observation spacing should also be chosen in order to
minimize the residual covariance between the orbital solution for
long-period planets and the solution for the parallax and the
proper motion of the primary; finally, $c)$ in the case of
observations of the known candidate planetary systems with one or
more planets, it might be necessary to time some observations on
the basis of the specific orbital phases, particularly in order to sample
adequately the pericenter passage of planets with highly eccentric
orbits. The primary focus of the next Section of this paper
is to discuss the SIM discovery space in terms of the smallest mass
the instrument might be capable to detect around the nearest 
stars. To this aim, in this Section we attempt to identify the
optimum timing of observations required in order to guarantee the
performance discussed in point $a)$. We pursue this objective
by analyzing a sample of possible options for the spacing of SIM
observations during the 5-yr mission lifetime, and neglecting for
the moment further investigations on the time interval between
pairs of quasi-orthogonal observations, randomly chosen in a range
of a few days. The optimal time separation will most likely depend
on SIM noise model, as well as overheads and other constraints
(occultations, exclusion angles), but such issues are not
addressed in this paper.

Setting $\sigma_\mathrm{d} = 2$ $\mu$as, we ran simulations with 
200 uniformly distributed star-planet
systems chosen to produce scaled signals in the range $1\leq S\leq 10$, 
for different choices of the orbital period $T$, which was allowed 
to vary in the range 5 days --- 5 years, and averaging over other 
orbital parameters. We used three reference stars per target and 
distributed $N_o = 24$ full 
observations over 5 years using a number of different spacings:
a) constant with 0.2-yr intervals; b) random uniform; c) proportional 
to the square of the number of observations 
($(3.15/365.25)\times n_\mathrm{obs}^2$, 
where $n_\mathrm{obs} = 1,\dots,N_o$); 
d) geometric series ($1.365^{n_\mathrm{obs}}/365.25$); 
e) logarithmic distribution 
($\mathrm{Exp}(\log 2 + \log 900\times(n_\mathrm{obs}/N_o))/365.25$).
We performed basic detection analysis via a standard $\chi^2$
test, with confidence level set as usual at 95\%, and estimated rms
errors on the orbital parameters, for each of the cases mentioned above.
The above choices for the time spacing of the observations are only 
illustrative, and by no 
means intended to be fully exhaustive, but can be regarded as a 
useful reference point when considering more sophisticated approaches 
to this important issue.

The probability of detection is more or less sensitive to the different 
timing of observations, depending on the value of the orbital period. 
For $T$ in the range between a few days and $\sim 1.5$ yr, all 
distributions are essentially equivalent, and systems with $S\geq 2.2$ 
are detected with high confidence (95\%), regardless of the choice of 
either of the above spacings. For $1.5\leq T\leq 5$ yr, the 
logarithmic and geometric series distributions are less favorable: for 
example, for $T = 2$ yr and $T = 5$ yr, a scaled signal 
$S\simeq 3$ and $S\simeq 4$ is required, respectively, in order to 
reach the $95\%$ detection probability threshold. In fact, with these 
two choices of time spacing $\sim 80-90\%$ of the observations occur within 
the first 1.5 years, and longer periods suffer from increasingly worse 
sampling. 
If $S\leq 2.2$, then different options for the timing of observations do not
improve detectability, regardless of the planet's orbital period. 
Finally we note that, in the case of equal
spacing, aliasing effects are found for periods that
happen to be shorter as well as exact integer fractions of the given 
sampling interval. 
In such cases, detection probability drops to zero, even for
$S\geq 2.2$, as sampling the orbit always at the same point translates in 
a small additive constant term in the observation equations 
that is completely absorbed in the least squares solution for the proper 
motion.

The situation is somewhat different when we focus on the accuracy in
orbit and mass determination. In particular, observations distributed 
according to the logarithmic sequence or the geometric series 
are preferable for measuring the orbital 
elements of planets in a few-days orbits, with an accuracy improvement 
of roughly 10-20\% with respect to the other distributions. In fact, for 
these two time spacings $\sim 40-50\%$ of the observations fall within 
the first 1-2 months of observing. 
For an intermediate period range between about 1 month and 1.5 yr, 
the accuracy in orbit reconstruction is comparable for all the different 
sampling distributions adopted. When $T$ approaches 5 years, again 
because of the lack of proper sampling at longer periods, the 
logarithmic and geometric distributions are less effective with respect 
to the other timing sequences, and orbital elements suffer a typical 
loss in accuracy of $\sim 10-20\%$. Within the limits of our observational 
scenario, the above behavior is essentially independent on the 
value of $S$. 
Finally, in the case of constant spacing, as we had already seen in
Section~\ref{estimate}, orbits with periods shorter than the
sampling time interval cannot be reconstructed, even when the signal
can be easily detected in the observation residuals to the
single-star fit.

Orbital eccentricity and inclination are expected to have a non-negligible 
impact on both detectability and measurability of planets. For example, 
a significant
degradation in accuracy can be expected for eccentric orbits with
periods longer than the mission lifetime (see for example 
Lattanzi et al.~\cite{latt00a}), due to the fact that deviations from
linear motion for a given target may go unrecognized if its orbit
around the system barycenter is not sampled in correspondence of the
pericenter passage. In the above analysis we find that the values of $e$ 
and $i$ are not critical in this scenario, as at least one entire 
orbit is sampled during the 5 years of simulated observations. 

For the purpose of this work, we have not 
analyzed the performance of different
orbital spacings for $T\geq 5$ yr, but rather focused on the range of
periods over which SIM observations are likely to provide best results.
Within the limited scope of this analysis, the relevant results are 
that distributing the observations according to a logarithmic or 
geometric series enhances accurate measurements of orbital parameters and 
masses for orbits of order of a few days, but degrades the probability 
of detection 
and orbital elements and mass determination as $T$ approaches the mission 
lifetime. Instead, a random, uniform distribution or 
a sequence proportional to the square of the number of observations, 
less likely to be subject to aliasing effects as in the case of 
equal sampling intervals, 
would be the preferred choice in order to achieve the best detection 
sensitivity and provide more accurate estimates of orbital parameters 
and masses over orbital periods ranging from $\sim 1$ month to the 
mission duration.

\subsubsection{Structure of the Standard Visit for
faint targets}\label{faint}

The observing sequence outlined in Section~\ref{scenario} and
utilized as a template to derive the results reported in 
Sections~\ref{detect} and~\ref{estimate} was only appropriate to 
bright targets and reference objects ($V\leq 11$). If the target
is fainter than $V = 11$, the structure of the Standard Visit should 
be modified. Within the constraint given by the 1-hr
performance specification period, many possibilities can
be investigated, with the primary aim of minimizing the error
$\sigma_\mathrm{d}$ on differential delay measurements. We consider 
two different options which meet the 1-hr constraint. The first 
consists of extending the integration time on the
target during each elemental fringe measurement, up to a maximum
of 3.5 minutes, which ensures the time for a target/reference star
elemental fringe measurement pair does not exceed 5 minutes
including overheads. A possible scenario for a $V = 16$ target
consists of observing blocks lasting $5\times n$ minutes composed
of $2\times 30\times n$ sec from overheads, $3.5\times n$ minutes
of integration on the target and $30\times n$ sec of integration
on reference stars, which we still assume bright ($V\leq 11$). In
this case, the Standard Visit accuracy per relative delay is
$\sigma_\mathrm{d} = 10.4/\sqrt{N_b}$ $\mu$as, driven by the
photon error on the target of 10 $\mu$as in each 3.5-minute
elemental fringe measurement (see Equation~\ref{differential}). In
order not to exceed the 1-hr performance specification period,
then for $n=3$ or $n=6$ reference stars we must require a maximum
of $N_b = 4$, or $N_b = 2$, respectively (and consequently
$\sigma_\mathrm{d} = 5.2$ $\mu$as or $\sigma_\mathrm{d} = 7.35$
$\mu$as). On the other hand, one can conceivably think of
increasing the number of observing blocks, keeping the duration of
each observing block within $2\times n$ minutes, as discussed in
Section~\ref{scenario}. A 30-sec integration on both the (faint)
target and the (bright) reference objects translates then in a
Standard Visit accuracy $\sigma_\mathrm{d} = 26.3/\sqrt{N_b}$
$\mu$as, and for $n=3$ and $n=6$ the limits on the maximum number
of observing blocks are $N_b=10$ and $N_b=3$, respectively (and
consequently $\sigma_\mathrm{d} = 8.28$ $\mu$as or
$\sigma_\mathrm{d} = 15.1$ $\mu$as). The first scenario is the
more efficient, as it allows for the presence of a possibly large
number of reference stars (to be preferred for the reasons
explained in Section~\ref{refer}), yet maintaining a higher
Standard Visit accuracy even with respect to the best-case
scenario for the second observing sequence.

For the purpose of the analysis carried out in the next Section,
we use two template observing strategies that take
advantage of the  observing scenarios discussed and results 
presented in Sections~\ref{scenario}, \ref{refer}, \ref{timing},
and above. In particular, we include 6 bright (all $V\leq
11$), astrometrically clean reference objects per target, and
distribute 24 epochs of 2 orthogonal Standard Visits randomly, uniformly 
spaced in time over the 5-yr SIM mission lifetime. Finally, 
we adopt somewhat conservative scenarios for the
single-measurement error $\sigma_\mathrm{d}$: $a)$ if the target
is bright ($V\leq 11$), then each Standard Visit is composed of
two unit observing blocks in which are executed 30-sec elemental
fringe measurements for target and reference stars, translating on
a Standard Visit accuracy $\sigma_\mathrm{d}\simeq 3$ $\mu$as per
each relative delay measurement; $b)$ for faint targets, say $V =
16$, the Standard Visit is composed of a single observing block in
which are executed 3.5-minute and 30-sec elemental fringe
measurements for target and reference stars, respectively, and
consequently the single-measurement error on each relative delay is
$\sigma_\mathrm{d} = 10.4$ $\mu$as. 
In both cases, a Standard Visit lasts approximately half an hour.

\subsection{SIM Discovery Space}\label{space}

In Section~\ref{estimate} we have shown how SIM's high-precision
astrometric measurements will provide information of great value
for the classification of planetary systems and overall
assessment of competing theories of planet formation and
evolution. In particular, in Section~\ref{known} we have illustrated
how extra-solar planets discovered by spectroscopic surveys would
fall within SIM's boundaries for accurate planet detection and
measurement, and discussed how they would undoubtedly constitute
an important laboratory for SIM. In fact, to go beyond a simple
catalogue of extra-solar planets, classification will
have to be made on the basis of the knowledge of their true
masses, shape and alignment of the orbits, structure and
composition of the atmospheres. The dependence of planetary
frequencies with age and metallicity will have to be understood.
Finally, important issues on planetary systems evolution, such as
coplanarity and long-term stability, will have to be addressed. 
However, the big picture will not be complete without crucial new
discoveries. The existence of giant planets orbiting on
Jupiter-{\it like} orbits (4-5 AU, or more), will have to be
established. Such objects, when found in systems harboring no
close-in giant planets, are the signposts for the discovery of
rocky planets in orbits closer to their parent stars,
maybe even inside the star's Habitable Zone, where water is liquid.

We currently have no
information about the existence of rocky planets orbiting any
other stars, except the rare pulsars~\citep{wols92}. 
SIM's exquisite astrometric precision will provide the 
opportunity for detection of planets with a range of masses
down to the mass of the Earth around the nearest stars. Answering
the age-old question of the uniqueness of our planet as a habitat
for life is clearly one of the highest priority objectives of
extra-solar planetary science, and the SIM measurements will 
uniquely complement the expectations coming from other ongoing and
planned planet-search surveys, for ground-breaking 
science in the field of formation and evolution of planetary systems. 

We have characterized the limiting performance of SIM in terms of 
its ability to detect (at the 95\% confidence level) Earth-class 
planets around a given star, as a function of the number of observations. 
As for the case of physical studies of
planetary systems discussed in Section~\ref{known}, also here
there exists a natural sample of potential targets for
observations from which SIM can extract important results: this
includes the nearest stars, around which SIM
could find planets as small as Earth, if they are present. 
The nearest stars are significant because their planets are easier
to detect by SIM and observe later by new telescopes that can
isolate and study their light. Such discoveries would provide
prime targets for the Terrestrial Planet Finder (TPF) to
characterize spectroscopically in terms of the potential for life.

Our nearest stellar neighbors fall into two main categories for
the purpose of SIM observations. The first category consists of
the few relatively luminous stars of spectral types K and earlier
($M_\star > 0.6$ M$_\odot$). The second category holds the many
low-mass, low-luminosity M dwarfs ($M_\star < 0.6$ M$_\odot$). The
latter are more astrometrically responsive to planets of a given
mass; the former includes stars more like the Sun and offers the
best opportunity to find habitable planets. 

As we have seen in Section~\ref{detect}, we can express the
mission detection sensitivity in terms of the minimum astrometric
signature for discovery $\alpha_{\mathrm{min}}$, defined by the
95\% confidence of detection. This quantity depends on the
Standard Visit accuracy $\sigma_{\mathrm{d}}$, and within our
simplified but realistic observational scenario it scales with the
square root of the number of measurement epochs $N_{o}$, and of
the number of reference stars $N_{r}$ (see Section~\ref{refer}).
For periods shorter than the mission duration, we found
$\alpha_{\mathrm{min}} = 2.2\,\sigma_{\mathrm{d}}$.

For any particular star, the boundary of discovery space---the
dividing line between detectable and non-detectable---is found by
equating $\alpha_{\mathrm{min}}$ to the astrometric signature of
the planet, defined by Equation~\ref{signature}. For a given
stellar mass and distance of the system, it is then possible to
determine the minimum detectable mass M$_{\mathrm{p,min}}$ as a
function of the semi-major axis of the planetary orbit
$a_{\mathrm{p}}$. To take into account the loss in sensitivity for
periods longer than the mission length $L$ (set to 5 yr), we have
parameterized M$_{\mathrm{p,min}}$ as follows:

\[
\mathrm{M}_{\mathrm{p,min}} = \left\{\begin{array}{ll}
2.2\,\frac{\displaystyle M_\star D}{\displaystyle a_\mathrm{p}}
\times\sigma_{\mathrm{d}} \times\sqrt{\frac{\displaystyle
24}{\displaystyle N_{o}}}\times \sqrt{\frac{\displaystyle
3}{\displaystyle N_{r}}} & \;\mbox{for\, $T \leq \frac{7}{9}L$}\\
 2.2\,\frac{\displaystyle M_\star D}
{\displaystyle a_\mathrm{p}}\times \sigma_{\mathrm{d}}
\times\sqrt{\frac{\displaystyle 24}{\displaystyle N_{o}}}
\times\sqrt{\frac{\displaystyle 3}{\displaystyle N_{r}}}\times
\sin^{-2}\left(\frac{\displaystyle \pi L}{\displaystyle 2T}\right)
& \;\mbox{for\, $T > \frac{7}{9}L$}
\end{array}
\right.
\]
In the plane defined by the mass of the planet M$_\mathrm{p}$ and
the orbital semi-major axis $a_\mathrm{p}$, this parametric
equation identifies a family of curves with the same shape, having
an absolute minimum at the value of $a_\mathrm{p}$ corresponding
to a period $T\simeq 4$ years, where the sensitivity is greatest.
The location of the minimum in semi-major axis increases linearly
with the distance and with the cube root of the stellar mass, 
following Kepler's third law.

To illustrate the potential of SIM for detection of low-mass
planets in the vicinity of the solar system, we have selected a
sample of 50 nearby stars within 10 pc from the Sun, divided into two
sub-samples of 25 stars each 
belonging to either of the two categories discussed
above. As general selection criteria, we have chosen to avoid
evolved systems and close binaries (with separations $<$ 10 AU).

For the more massive stars (M$_\star > 0.6$ M$_\odot$), the main
goal is to detect Earth-class planets in the Habitable Zone (HZ).
According to the conventional wisdom, a habitable Earth must orbit
at a distance from its star where liquid water is stable on its
surface. In its classic definition, the 
inner boundary of the HZ~\citep{kasting93} is
located at the distance from the star at which a runaway
greenhouse effect is generated, which induces water loss
via photolysis and hydrogen loss; the outer boundary is located at 
the distance from the star at which $CO_2$ clouds start increasing
the planet albedo in a way to cool the surface down to the point
of freezing water. For low-mass stars this region is very narrow
and located at distances much less than 1 AU, while it is wider
and located at distances much greater than 1 AU for high-mass
stars. The boundaries of this region change in time due to the evolution
of the central star, and the concept of Continuously 
Habitable Zone (CHZ) must be introduced, to identify the region of
space around a given star which can be considered habitable at
different times. Furthermore, in recent 
works~\citep{forget97,kasting98,forget00} it has been argued that 
either $CO_2$ clouds or the presence of abiotic and/or biogenic 
$CH_4$ in the atmosphere should tend to warm a planet's surface, thus HZs 
might be significantly wider than previously thought. For the 
purpose of this investigation, we 
define operationally the center of the HZ in terms of the orbital 
period $T_{HZ}$ (in years) and stellar mass M$_\star$ (in solar masses):
\begin{equation}
\frac{T_{HZ}}{T_\oplus} =
\left(\frac{M_\star}{M_\odot}\right)^{7/4}
\end{equation}
This formula, in which $T_\oplus$ is the Earth's orbital period, 
roughly holds for Main Sequence stars of spectral type F through K. 
The inner edge of the HZ is located at $T_{HZ,i} \simeq 0.7\,
T_{HZ}$, and the outer edge at $T_{HZ,o} \simeq 2\, T_{HZ}$.

If a star is not of spectral type K through F, then the
characteristics of the HZ change dramatically: very
bright and hot stars (spectral type A through O) last far too
short a period of time on the Main Sequence (up to a few millions
years) to allow for the development of complex life forms 
(according to the typical time-scales of biological evolution on
Earth); M dwarfs instead are thought to be non-ideal environments for
harboring a complex biology for two different reasons: 
$a)$ the HZ of a low-mass star is well within the tidal locking 
radius~\citep{kasting93}. 
Whenever synchronous rotation of the star and planet 
is established, life may be hampered by condensation of the atmosphere 
on the perpetually cold dark side of the planet; $b)$ large 
stellar flares are common in M-type dwarfs, and they would tend 
to sterilize life on a regular basis.

Nevertheless, if arguments can be raised against the intrinsic
relevance of the HZs of M dwarfs as potential life-sustaining 
environments, stars of late spectral type are
specially interesting because of the more favorable planet/star
mass ratio, which allows easier detection of low-mass planets. 
Furthermore, due to the present results from planet searches 
being biased towards 
solar-type stars, issues such as the dependence of planetary 
frequencies with the spectral type still need to be addressed.

For each of the two stellar sub-samples discussed above, we have set a 
mass-sensitivity threshold, and determined, within the framework 
of the two template observing strategies outlined in Section~\ref{faint}, 
the number of observations needed to reach it. In particular: $1)$ 
for the sub-sample with M$_\star > 0.6$ M$_\odot$, we have set the 
mass-threshold for detection to 3 M$_\oplus$ at the 
center of the HZ, except for the
case of $\alpha$ Cen A-B, for which, given their proximity to the Sun, 
we set the goal to be a 1-M$_\oplus$ sensitivity at the center of the
HZ; $2)$ for the sub-sample of stars with 
M$_\star < 0.6$ M$_\odot$, we have set the same threshold to 
1 M$_\oplus$ at the most sensitive point of the discovery-space 
curve, wherever it occurs for each star in semi-major axis 
(which always corresponds to about a 4-year period).

In Figure~\ref{fig9} we have plotted the parametric equation defining
the minimum detectable mass by SIM as a function of semi-major
axis for the 10 solar-type stars and for the 14 M dwarfs that require 
the lowest amount of observations in order to reach the respective goals. 
The curves relative to each star are color-coded by the number
of observations needed to achieve the requested sensitivity. As it
can be easily seen, for all the stars in the solar-type sub-sample additional
observations are needed, up to about 6 times the default $N_{o}
= 24$, in the case of $\eta$ Cas. 
Instead, we find that 24 observations are sufficient to 
{\it exceed} the goal for all of the M dwarfs plotted in Figure~\ref{fig9}. 
As expected, the
highest sensitivity is reached in the case of Proxima Cen, where
the template observing strategy for a faint target outlined in
Section~\ref{faint} allows for a minimum detectable mass of 
$\sim 0.2$ M$_\oplus$.

These results have been obtained in the context of a somewhat conservative 
choice for the Standard Visit accuracy for both bright ($V\leq 11$) and 
faint targets (see Section~\ref{faint}), and assuming bright reference 
stars. As a consequence, 
the amount of full two-dimensional observations needed to 
achieve the goals of detection of Earth-mass planets in the HZ of 
solar-type stars within 10 pc from our Sun is very large. Instead, very 
low-mass planets might be revealed orbiting nearby M dwarfs with more 
relaxed requests in terms of SIM observing time and single-measurement 
accuracy. 

Our findings are illustrative of some of the many important issues 
future observing programs with SIM operated 
in narrow-angle mode will need to debate, at the moment of the final 
selection of targets. In particular: $(a)$ an optimal tradeoff between the
number of stars surveyed and the depth of the search will have to be 
established; $(b)$ the details of the adopted strategies for observing 
will have to be refined to maximize the ratio of the number and intrinsic 
scientific interest of the objects in a target list to the fraction of the 
total observing time utilized; $(c)$ the specific merit of any particular 
star itself will have to be discussed, which has both scientific
facets (type of star, theories of planet formation) and technical
aspects (availability of a robust, bright reference frame, properties of 
astrometric noise of the target).

\section{Summary and Conclusions}

Since the establishment of the existence of the first extra-solar
planet orbiting a solar-type star~\citep{mayor95}, 
the approach to sciences of stars
and planets has dramatically changed. Now, answers are sought to
more advanced questions about the formation and evolution of
planetary systems and the existence of rocky, perhaps habitable
planets.

Precision astrometry constitutes a fundamental complement to other
search techniques. Today, monolithic telescopes and optical
interferometers are being built or designed, which will provide 
accurate astrometric measurements, both from 
ground~\citep{mariotti98,booth99,colavita99} and in 
space~\citep{danner99,roser99,perryman01}.

In this paper we have used extensive end-to-end numerical
simulations of narrow-angle astrometric measurements with the
Space Interferometry Mission and the subsequent statistical
analysis of the simulated dataset in order to quantify the
potential of SIM for the discovery and characterization of single
planets around single stars in the vicinity of the solar system.
Utilizing a simplified, but realistic, error model for SIM operated in
narrow-angle mode, and adopting a reasonable, flexible template observing
scenario (Sections~\ref{scenario}), we have:
$a)$ defined the boundaries for secure planet
detection and accurate determination of orbital elements and masses, as
function of the basic SIM capabilities and properties of the
observed systems (Sections~\ref{detect},~\ref{errors}, and~\ref{known}), 
$b)$ evaluated the impact of different observing strategies on the 
boundaries for detection and orbit reconstruction (Sections~\ref{refer} 
and~\ref{timing}); $c)$ adopting template observing strategies for 
both bright ($V\leq 11$) and faint targets (Section~\ref{faint}), 
illustrated SIM discovery potential in terms of its ability to 
detect terrestrial planets around a sample of the closest stars
(Section~\ref{space}). Our main results can be summarized as follows.

\begin{itemize}
  \item[(1)] secure detection (at the 95\% confidence level) will
  be possible for planets producing an astrometric signature
  $\alpha_{\mathrm{min}}\sim 2.2$ times larger than the Standard Visit 
  accuracy $\sigma_\mathrm{d}$, for periods shorter than 5 years, 
  the nominal mission lifetime;
  \item [(2)] in the same period range, the mass of the planet and
  the full set of orbital elements will be determined with a typical
  accuracy of 20-30\% for objects producing a signal 
  $\sim 2\,\alpha_{\mathrm{min}}$; for mass and
  inclination measurements accurate to 10\%, the required signal
  is $\sim 10\,\alpha_{\mathrm{min}}$ and
  $\sim 15\,\alpha_{\mathrm{min}}$, respectively; analyzing how 
  the set of presently known extra-solar planets would fall within the 
  boundaries for reliable detection and accurate mass and orbit 
  determination, we 
  find that about 75\% will be detected and 50\% will have orbital
  elements and masses measured to 10\%, or better;
  \item[(3)] the detection threshold scales similarly with the number of 
  observations ($\sqrt{N_o}$) and reference stars ($\sqrt{N_r}$); 
  random uniform and geometric distributions of the observations are 
  preferred for achieving the best detection sensitivity and more 
  accurate estimates of orbital parameters and masses in the period range 
  between $\sim 1$ month and 5 yr, and $\leq 1$ month, respectively; 
  \item [(4)] due to the very small astrometric signature induced on 
  the parent star, reliable detection of Earth-class planets 
  in the Habitable Zone of
  the closest solar-type stars will be possible, but demanding in terms 
  of number of full observations per target and measurement precision; 
  instead, around 
  the nearest M dwarfs, more relaxed constraints on the number of 
  observations and single-measurement errors would still ensure 
  detection of planets as small as Earth.
\end{itemize}

Our findings indicate how SIM, with its unprecedented
astrometric precision, will be a valuable tool
for discovering planets around stars other than the Sun. Among the
new generation of instruments designed to study extra-solar
planets, SIM will be able to provide unique insights towards the
understanding of planetary systems in their generality and
investigating the habitability of other worlds than Earth. Today 
two factors hamper the transition from the present
cataloguing phase to the more fundamental classification phase,
where, for example, mass might be operationally used as one of the
genesis indicators which would help discriminate between
planets and brown dwarfs: $a)$ mass uncertainty
for the radial-velocity discoveries (due to inclination angle
ambiguity), and $b)$ incompleteness in the mass range corresponding to
solar system planets (due to inadequate sensitivity). By
determining the true rather than the projected orbit of the planet
(as with radial-velocity techniques), SIM measurements
will remove the inclination angle degeneracy and associated
companion-mass uncertainty for the presently existing planets, as
well as for those the instrument will discover directly. 
Furthermore, by ruling out the presence of Earth-mass planets
around the nearest stars, SIM will be capable of addressing for
the first time the role of rocky cores in the complex scenarios of
planetary formation and evolution, and start to investigate their
potential habitability. In fact, SIM astrometry will be important
in investigating the Habitable Zones of stars with known planets in
wide orbits: those systems in which the Habitable Zone and the
zone in which planet formation has not been disrupted by the
presence of the known giant planet overlap~\citep{wetherill96} would
immediately become high-priority targets for SIM narrow-angle 
observations, to 
search for terrestrial planets and find evidence of the existence
of planetary systems resembling our own.

Another crucial area in which SIM measurements might have a 
significant impact is the study of
multiple-planet systems: the remarkable pattern of
low-eccentricity orbits and coplanar structure of the solar system
are commonly thought to be fossil evidence of the planets having
accumulated in a dissipative protoplanetary 
disk~\citep{lissa93,pollack96}. The wide variety of planetary masses
and orbits found by radial velocity techniques have called into
question the generality of such ideas, suggesting that
significant orbital evolution may be needed to explain the
high-eccentricity orbits~\citep{arty92,weiden96,lin97,mazeh97} and 
the {\it Hot Jupiters} at very small orbital 
radii~\citep{lin96,murray98,lin00}. By answering the
seemingly simple question of whether multiple-planet orbits are
coplanar, SIM might confirm that some other planetary systems are
similar to our own and similarly indicative of origin in a
quiescent, flattened disk. Or, SIM measurements might provide
evidence that other systems are truly different, with large
relative orbital inclinations, which could point to either an
early, chaotic phase of orbital evolution or formation by another
mechanism such as disk 
instability~\citep{kuiper51,cameron78,boss97,boss00,bossetal02}. 
The simulation of SIM observations of 
extra-solar multiple-planet systems, the quantification of the 
instrument capability in discovering and measuring systems of planets,
as well as its ability in determining coplanarity of 
multiple-planet orbits, will constitute the core of the results presented 
in paper II.

\acknowledgements

Special thanks are due to G. Fritz Benedict, Alan Boss, George Gatewood, 
Todd Henry, David Latham, and Robert Reasenberg for lending initial 
impetus and support to this investigation. Over the course of this 
work, we have benefited from discussions with numerous colleagues, and 
especially Mike Shao, Steve Unwin, and Dave van Buren. We are also 
grateful to an anonymous referee for her/his helpful comments. This work has 
been carried out with the partial financial support of JPL under 
Contract 960506 to the Space Telescope Science Institute. A.S. and 
M.G.L. gratefully acknowledge support from the Italian Space Agency 
under Contract ASI-I/R/117/01.


\clearpage



\figcaption[]{Equal probability contours for planet detection at
95\% confidence, as a function of orbital period $T$ and
astrometric signature $\alpha$. Detection probability averaged
over all other orbital parameters. The dashed lines indicate the
equivalent signature at the given distance for a solar-mass
primary and for a 20 $M_\oplus$ planet (1 $M_\mathrm{J}$ in
parenthesis) \label{fig1}}

\figcaption[]{Isoprobability contours for various detection
probabilities as function of the orbital period $T$ and the
distance $D$, for a 1-M$_J$ planet orbiting a 1 M$_\odot$ star
\label{fig2}}

\figcaption[]{Detection probability as function of the ratio
between astrometric signature $\alpha$ and single-measurement
error $\sigma_\mathrm{d}$, for various orbital periods
\label{fig3}}

\figcaption[]{Orbital fits for 200 simulations of a 1-M$_J$ 
planet around a 1-M$_\odot$ star with a 1-yr (black histograms), 
5-yr (red histograms), and 12-yr (yellow histograms) period, at 
$D$ = 100 pc ($\alpha\simeq 10$ $\mu$as), 200 pc ($\alpha\simeq 15$ $\mu$as), 
and again 100 pc ($\alpha\simeq 50$ $\mu$as), respectively. 
The panels
present the distributions of the fitted values of the most
relevant parameters with respect to their ``true'' values.
Eccentricity, inclination, and phases are chosen randomly for each
simulated system. The 1-yr periodicity is best determined, while 
slightly larger errors on the parallax and semi-major axis are 
evidence for the (weak) coupling between orbital period and
parallactic motion \label{fig4}}

\figcaption[]{The motion on the sky of a system composed of a 
1-M$_\odot$ star and a 1-M$_J$ planet on a 5-yr period, at $D = 200$ 
pc, as `seen' by SIM during its 5-yr mission. In the two upper
panels the solid lines represent the true motion along the X- and
Y-axis, the triangles are the computed positions at the epochs of
SIM observations, after the single-star fit. The post-fit
residuals, as function of the time of observations, are shown as
diamonds in the relative sub-panels, clearly highlighting the
presence of the periodic perturbation due to the orbiting planet;
the lower left panel shows the combined apparent motion of the system 
on the sphere in two dimensions (solid line), and superposed the
fitted positions at the epochs of SIM observations (asterisks)
\label{fig5}}

\figcaption[]{The orbital motion of the central star in the system of 
Figure~\ref{fig5} around the common barycenter: 
the solid line in the upper two panels represents
the true orbit of the star projected on the X- and Y-axis,
respectively, while the squares are the computed positions at the
epochs of SIM observations, after the full Keplerian fit; in the
two sub-panels the post-fit observation residuals as a function of
the time of observations reveal do not reveal any additional
periodic behavior, and are consistent with the
measurement errors, providing confirmation of the high
accuracy in the reconstruction of the planetary orbit (as shown by
the rms errors on the fitted parameters in Figure~\ref{fig4}); the
two lower panels show the combined orbital motion of the star in two
dimensions (solid line), and superposed the epochs of SIM
observations after the single-star fit (crosses in the lower left
panels) and after the full orbital fit (asterisks in the lower
right panel), respectively \label{fig6}}
\figcaption[]{The boundaries of secure detection and accurate 
mass and orbital parameters determination compared to the known 
extra-solar planets, which are plotted for the
{\it minimum} case: orbit viewed edge-on, true mass equals
radial-velocity minimum mass, and astrometric signature minimum.
The radius of each planet's symbol is proportional to the cube
root of the minimum mass. As described in the text,
lines of different shape represent the minimum astrometric
signature for 95\% probability of detection (solid), the minimum
astrometric signature for 10\% accuracy measurements of the
(minimum) mass of the planet (dashed-dotted), and the minimum
astrometric signature for 10\% accuracy measurements of the
(maximum) inclination angle (dashed), respectively. The true
astrometric signature, which is proportional to the true mass,
will be generally higher---much higher in some cases---with the
effect that more reliable detections and orbital fits will be 
possible
\label{fig7}}

\figcaption[]{Detection horizon vs. number of two-dimensional
observations for 3 and 6 reference stars, and for a Jupiter-Sun
system with orbital period $T = 5$ years (near the peak in the
detectability curves in Figures~\ref{fig1} and~\ref{fig2}). 
A detection probability $P\geq 95\%$ is assumed
\label{fig8}}

\figcaption[]{Detectable mass as a function of semi-major axis for
a sample of stars within 10 pc from the Sun. Discoverable planets
lie above the solid curve for each target. Curves are color-coded
by number of observations, from yellow (minimum of 24 full observations 
with the template observing strategies outlined in Section~\ref{faint}) 
to green (maximum of
130 observations for $\eta$ Cassiopei\ae\,); the color scale is
shown in the lower-right corner. The minimum of each curve
corresponds to a period of 4 years, where the sensitivity is
greatest; the steep rise to the right illustrates the loss of
sensitivity as the period approaches and exceeds the mission
duration. The position of the vertex moves from the left (M
dwarfs) to the right as the stellar mass increases; for clarity,
only the vertex is shown for several M dwarfs. The curves for Gl
702 A and $\tau$ Ceti overlap; both labels are listed for one
curve (all curves have the same shape). The thick blue segment
shows the approximate Habitable Zone around each star. For
low-mass stars, the Habitable Zone falls to the left of the graph.
The diagonal purple line corresponds to best-case detectability by
radial-velocity searches (edge-on orbit) with an accuracy of 1
m/s, assuming $\mathrm{M}_\star = 1 \mathrm{M}_\odot$ \label{fig9}}

\clearpage
\begin{figure}
\plotone{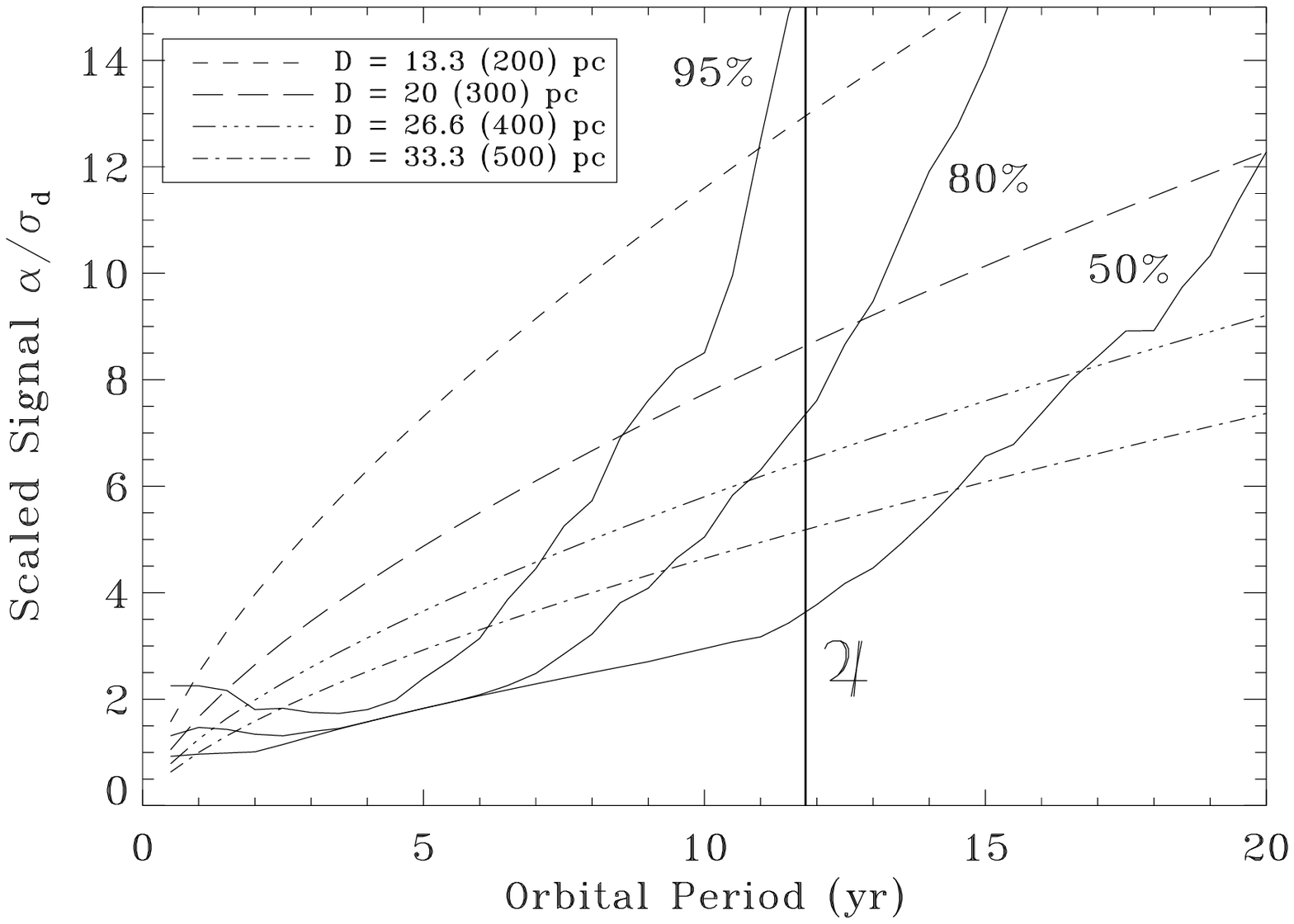}
\end{figure}
\clearpage
\begin{figure}
\plotone{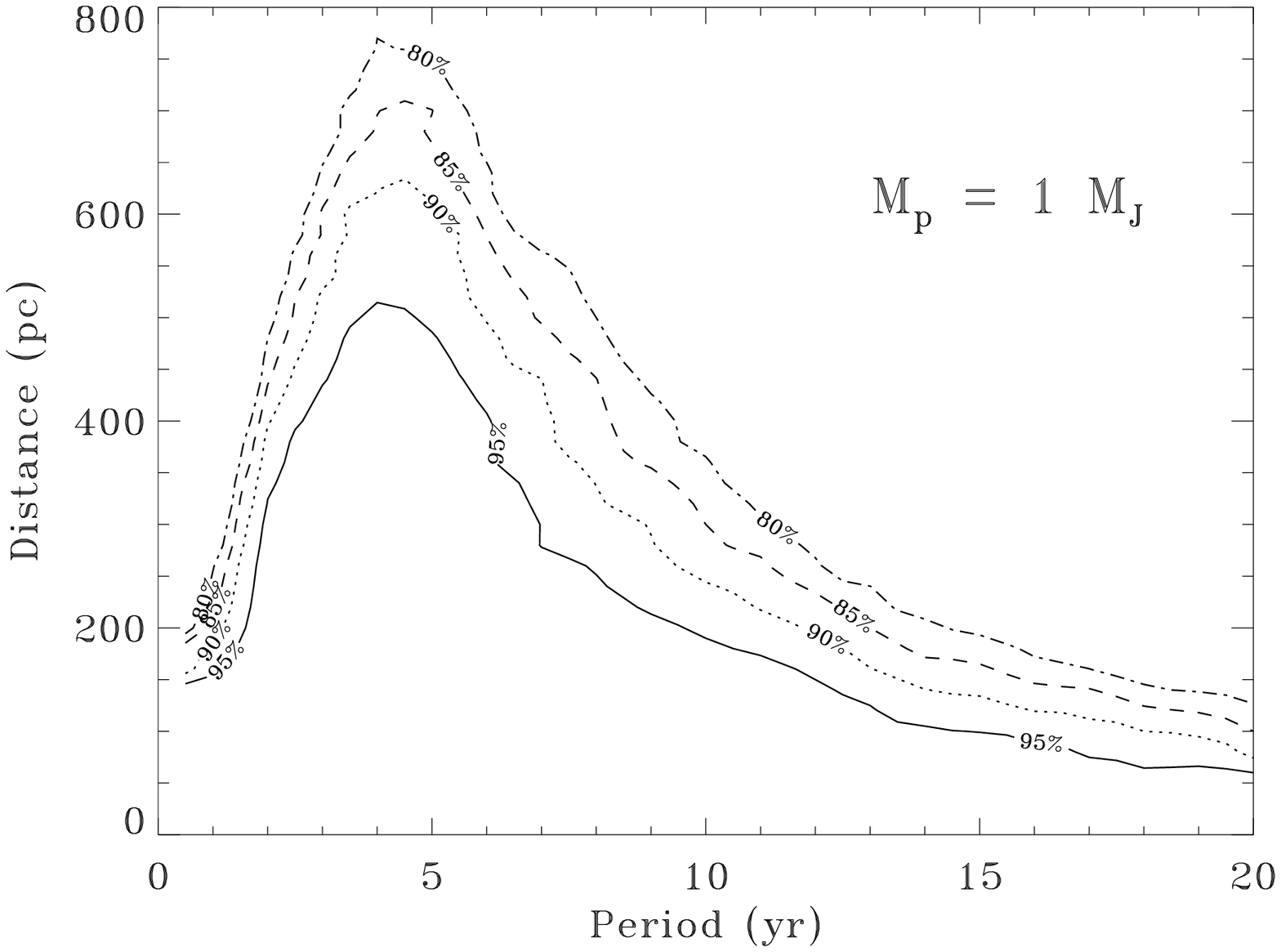}
\end{figure}
\clearpage
\begin{figure}
\plotone{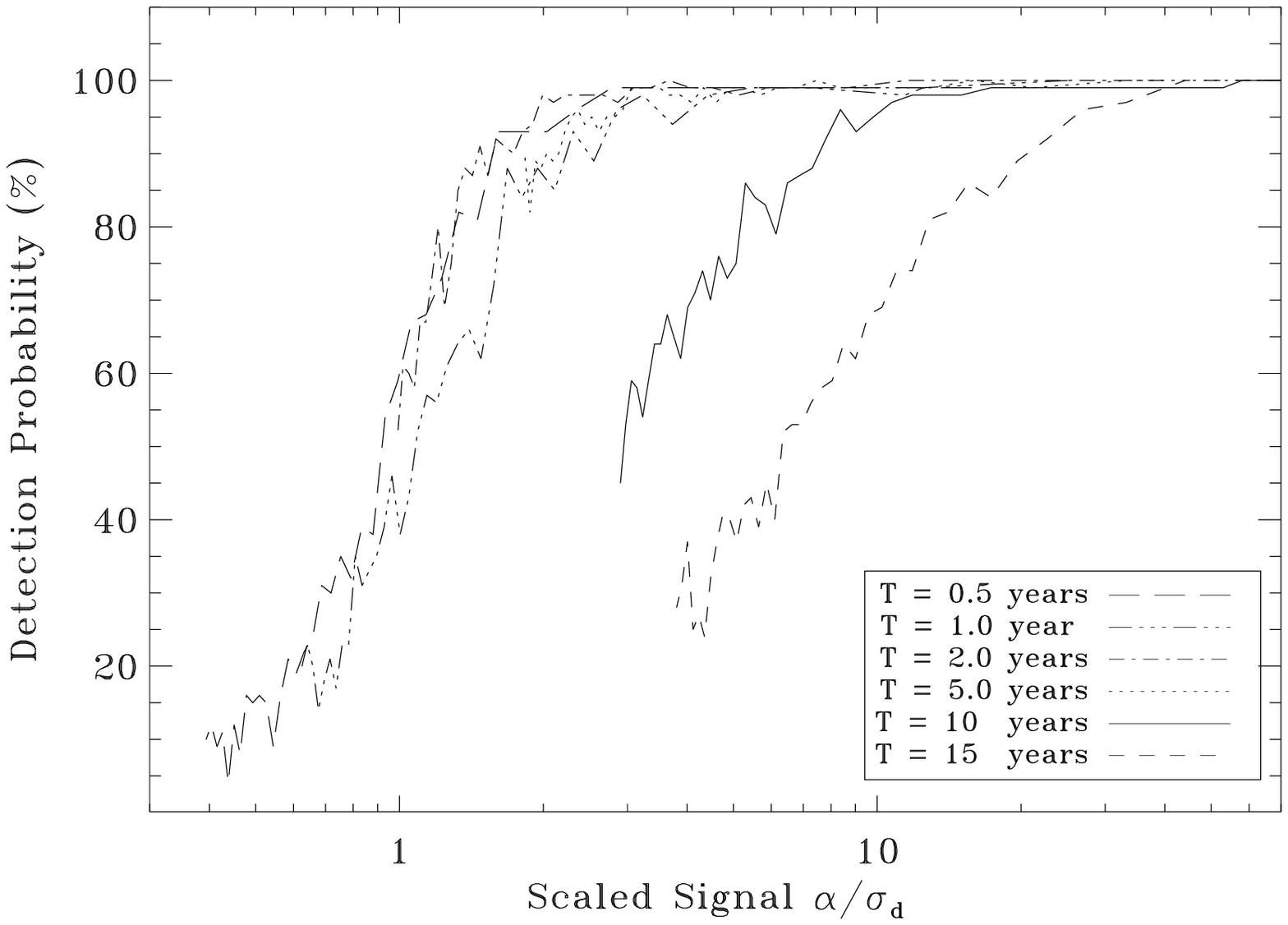}
\end{figure}
\clearpage
\begin{figure}
\includegraphics[bb=65 0 700 680,width=22cm,height=24.5cm]{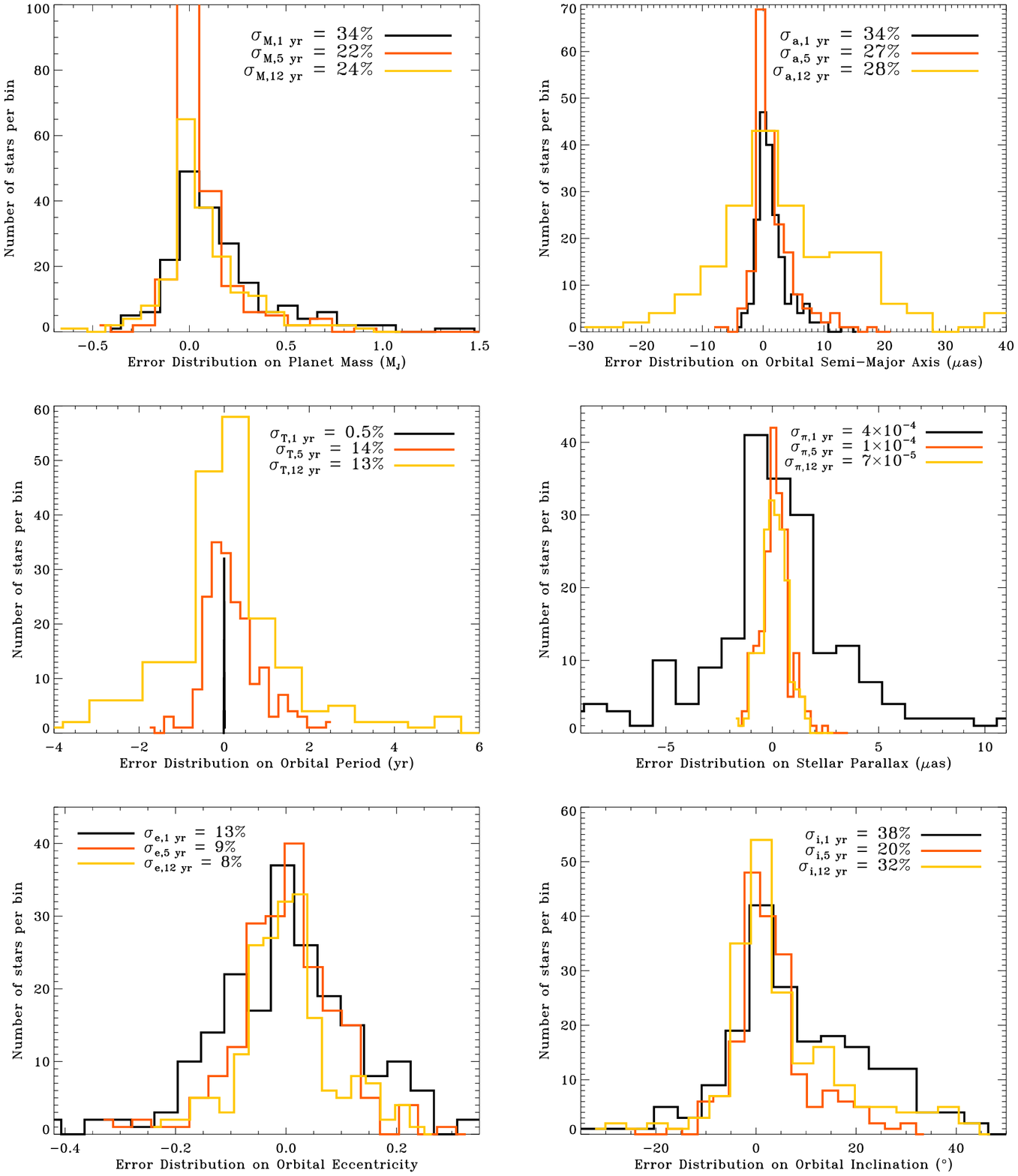}
\end{figure}
\clearpage
\begin{figure}
\plotone{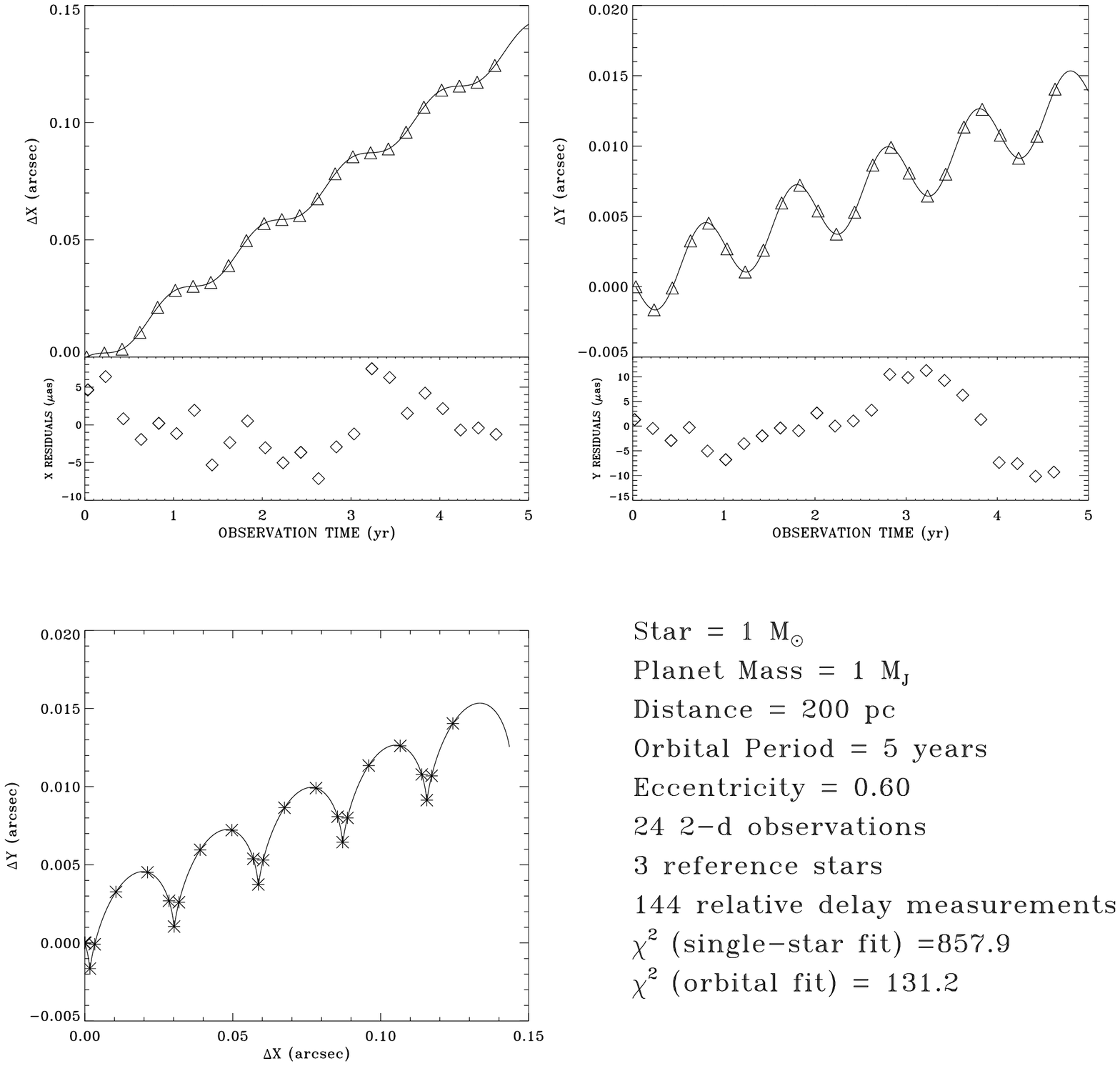}
\end{figure}
\clearpage
\begin{figure}
\plotone{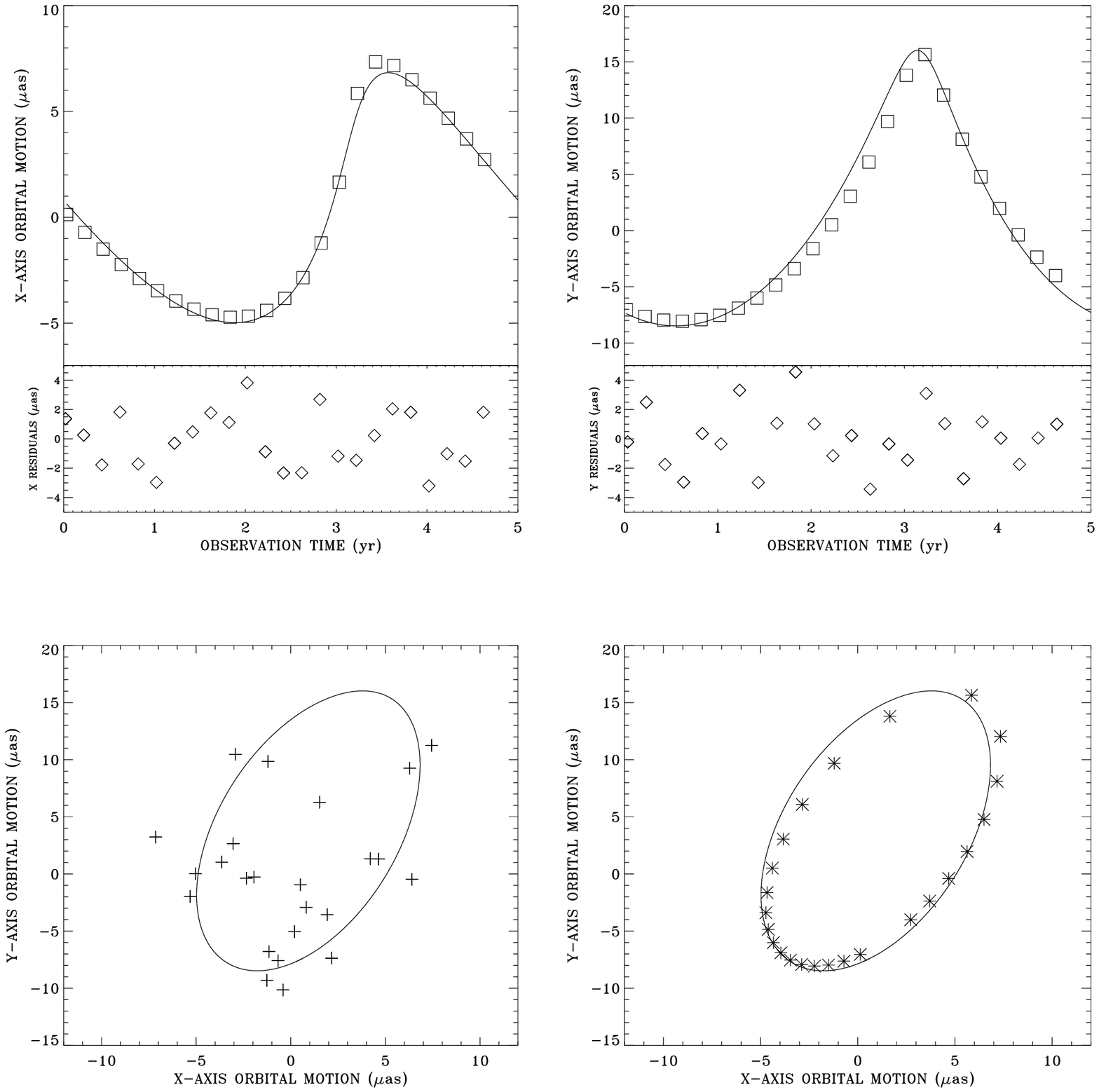}
\end{figure}
\clearpage
\begin{figure}
\includegraphics[bb=70 -140 609 637,width=19.cm]{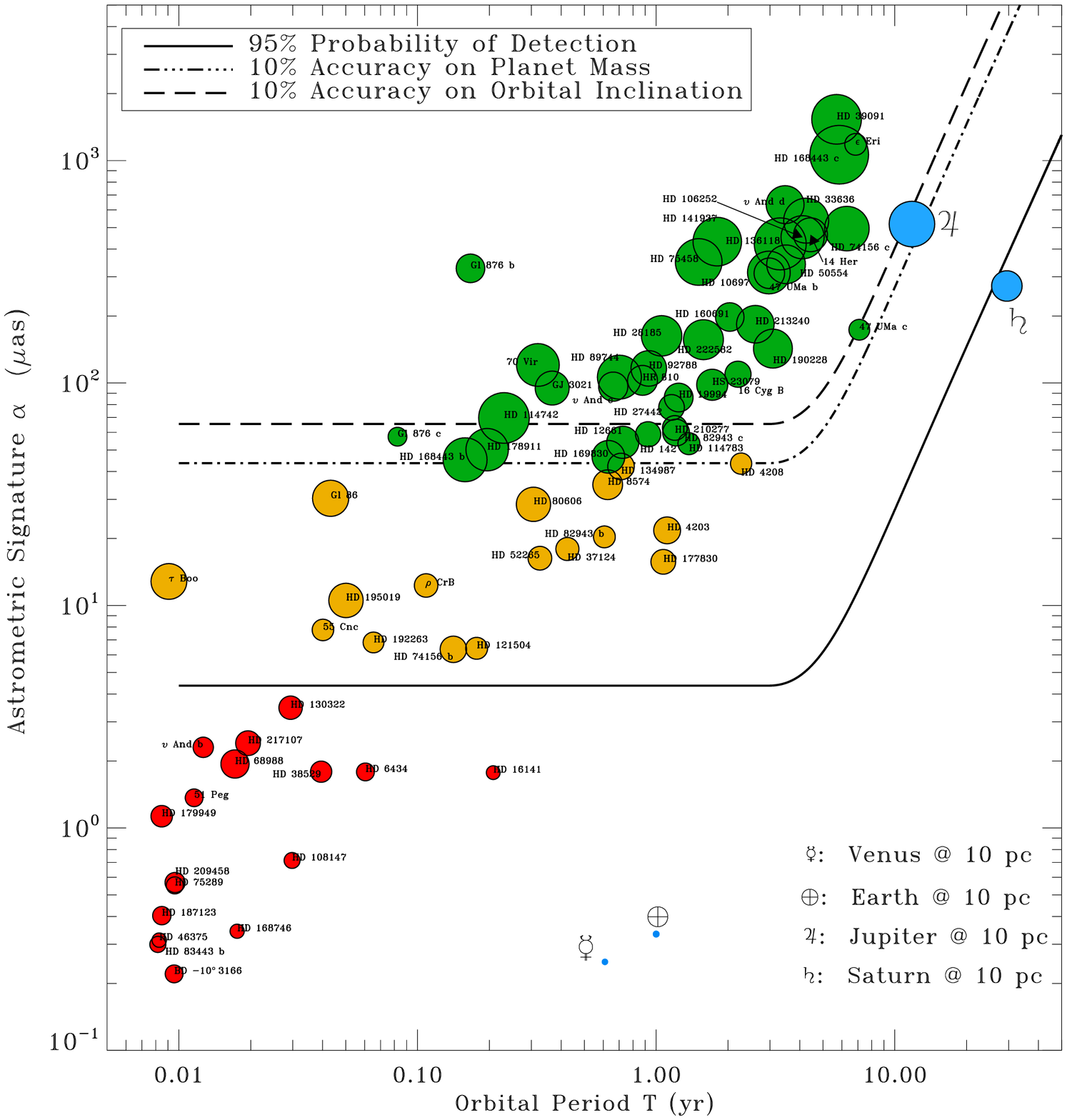}
\end{figure}
\clearpage
\begin{figure}
\plotone{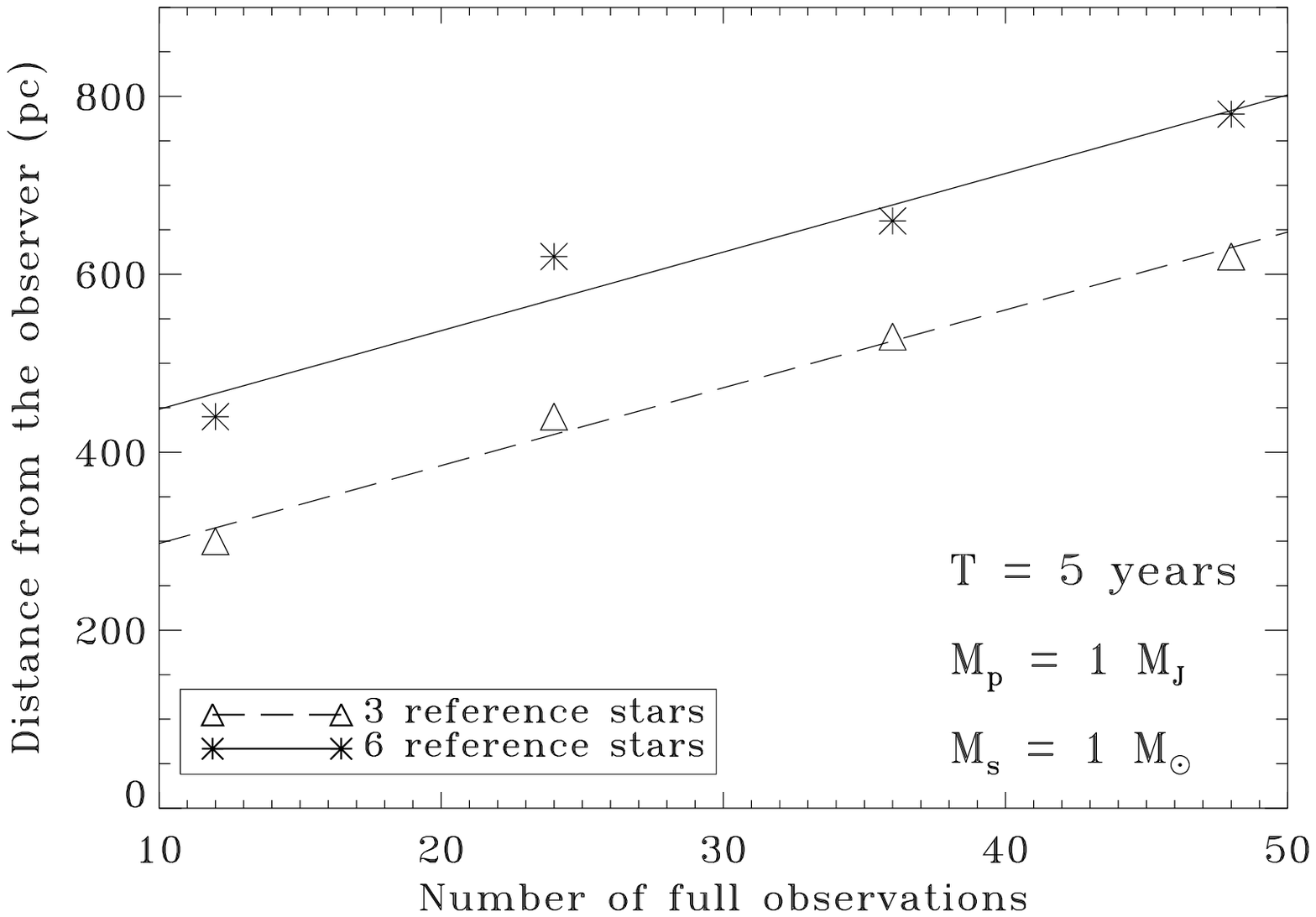}
\end{figure}
\clearpage
\begin{figure}
\begin{center}
\includegraphics[bb=90 -50 612 792,width=22cm]{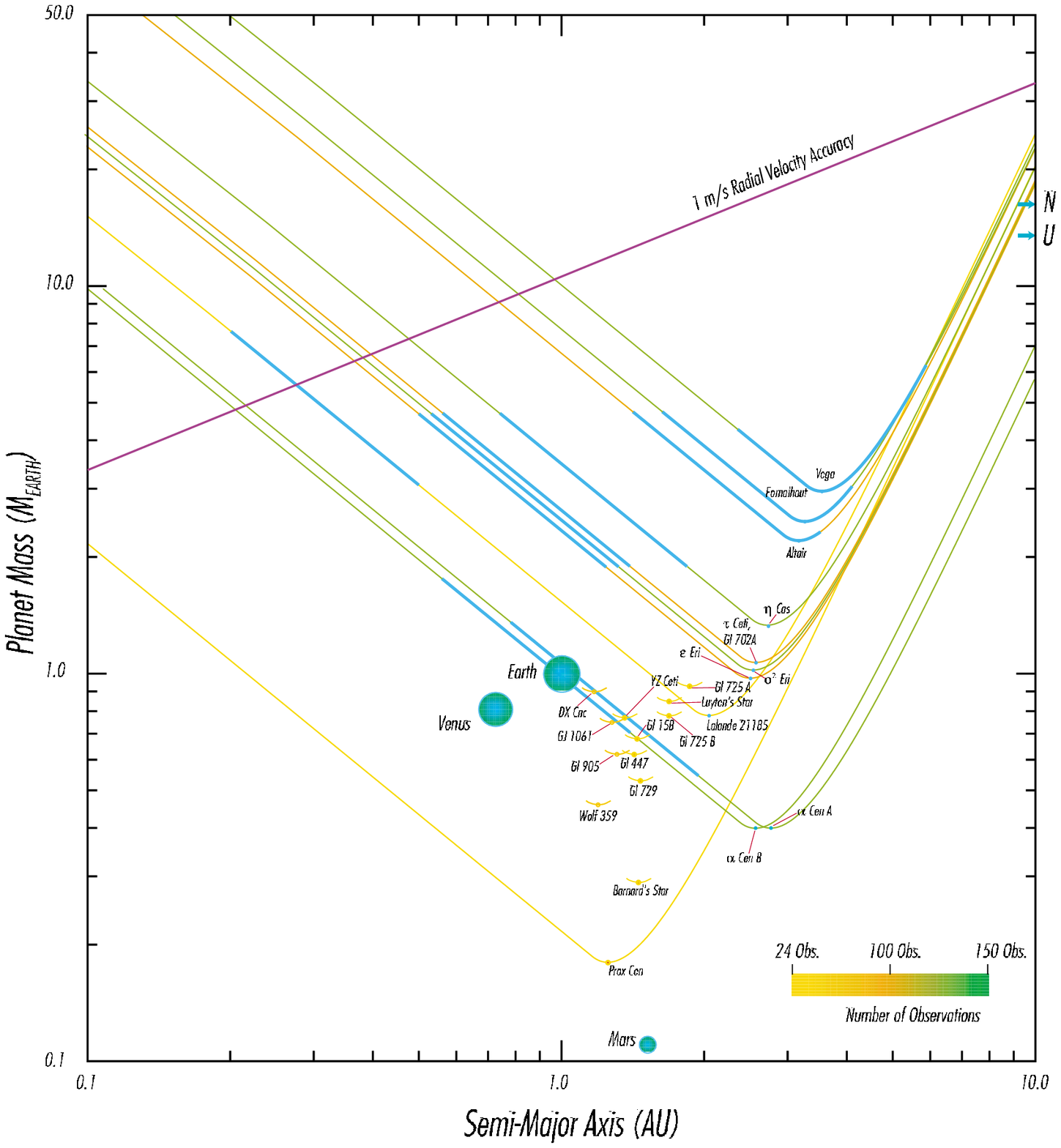}
\end{center}
\end{figure}


\end{document}